\documentclass[aps,onecolumn,nofootinbib,tightenlines,superscriptaddress]{revtex4}

\usepackage{amssymb,amstext,amsmath,amsthm,slashed}
\usepackage[dvips]{graphicx}
\usepackage{latexsym}
\usepackage{psfrag}
\usepackage{amsfonts}
\usepackage{color}
\usepackage{verbatim}
\usepackage{bbm}
\usepackage{dcolumn}
\usepackage{tabularx}
\usepackage{leftidx}
\usepackage{tensor}
\usepackage{boxedminipage}
\usepackage{fancyvrb}
\usepackage{float}
\usepackage{wrapfig}
\usepackage{booktabs}
\usepackage{multirow}
\usepackage{subfigure}
\usepackage{dsfont} 
\usepackage{mathrsfs}
\usepackage[hidelinks]{hyperref}
\usepackage{scalerel,stackengine}
\usepackage{amsthm}
\usepackage[dvipsnames]{xcolor}
\usepackage{natbib}

\setcitestyle{authoryear,round}

\theoremstyle{plain}

\newtheorem*{theorem*}{Theorem}

\newcommand{\1}{\mbox{\rm 1 \hspace{-1.05 em} 1}}

\begin{document}

\title{On Penrose's Analogy between Curved Spacetime Regions \\ and Optical Lenses}


\author{Dennis Lehmkuhl} 

\email{dennis.lehmkuhl@uni-bonn.de}

\affiliation{Lichtenberg Group for History and Philosophy of Physics - Faculty of Philosophy, University of Bonn, 53113 Bonn, Germany \vspace{0.25cm}}


\author{Christian R\"oken\vspace{0.25cm}} 

\email{croeken@uni-bonn.de}

\affiliation{Lichtenberg Group for History and Philosophy of Physics - Faculty of Philosophy, University of Bonn, 53113 Bonn, Germany \vspace{0.25cm}}

\affiliation{Department of Geometry and Topology - Faculty of Science, University of Granada, 18071 Granada, Spain \vspace{0.25cm}}


\author{Juliusz Doboszewski} 

\email{jdobosze@uni-bonn.de}

\affiliation{Lichtenberg Group for History and Philosophy of Physics - Faculty of Philosophy, University of Bonn, 53113 Bonn, Germany \vspace{0.25cm}}

\affiliation{Black Hole Initiative - Harvard University, Cambridge, MA 02138, USA \vspace{0.25cm}}


\date{October 2023 / June 2024}

\begin{abstract}
\vspace{0.4cm} \noindent \textbf{\footnotesize ABSTRACT.} 
\, We present a detailed analysis of Penrose's gravito-optical analogy between the focusing effects of particular families of Ricci- and Weyl-curved spacetime regions on the one hand, and anastigmatic and astigmatic optical lenses on the other. We put the analogy in its historical context, investigate its underlying assumptions, its range of validity, its proof of concept, and its application in Penrose's study of the notion of energy flux in general relativity. Finally, we examine the analogy within the framework of Norton's material theory of induction.
\end{abstract}

\setcounter{tocdepth}{2}

\vspace{0.1cm}

\maketitle

\tableofcontents

\section{Scope of the Paper} \label{SectionI}

\noindent One of the most beautiful ideas stemming from 20th century physics is that what we feel as the force of gravity is actually the curvature of spacetime. Thus, the motion of the Earth around the Sun is often explained by analogy to a spacetime being like a sheet of cloth that is curved by the heaviness of the Sun, with the much lighter Earth moving on a geodesic of this curved spacetime, which we observe as the Earth moving on an ellipse around the Sun.

There is, however, a less well-known analogy concerned with spacetime curvature. This analogy was introduced by Roger Penrose in the 1960s, and it has as its purpose a.) to better understand two particular types of spacetime curvature, namely Ricci curvature and Weyl curvature, and b.) to come to grips with the notion of energy flux in the context of general relativity. Penrose imagines a beam of light rays traversing a region of spacetime, and wonders what effects the presence of Ricci curvature and Weyl curvature, respectively, will have on the beam. He then compares this with light traversing different types of elementary lenses in geometrical optics. Penrose's idea is that there is an analogy between the effects of Ricci-curved and Weyl-curved regions of spacetime on the one hand, and the effects of anastigmatic and astigmatic lenses on the other.\footnote{Penrose's analogy thus connects the influence that gravitational fields have on light according to general relativity to the influence that optical lenses have on light according to geometrical optics. An even earlier connection between a theory of gravity and optics predates the connection forged between gravity and spacetime curvature in general relativity. In Einstein's 1911 paper on static gravitational fields, which came out two years before his move of identifying the gravitational potential with the metric tensor, he found that, given the equivalence principle, the speed of light differs from point to point when light is moving through a static gravitational field. This allowed Einstein to transfer Huygens' principle from wave optics, where it used to derive the angle at which a beam of light is refracted when it enters an optical lens, to his nascent gravitational theory, where it enabled him to predict the angle at which a beam of light passing through the gravitational field of the Sun would be refracted (though the angle he predicted was only half of what he later predicted with the finalized theory of general relativity in hand); see \citet{Einstein:1911h}, Section 4. We are grateful to an anonymous referee for pointing us to this earlier transfer from optics (here wave optics) to gravitational theory, though it should be noted that in contrast to Penrose, Einstein does not explicitly speak of an analogy between gravitational fields and optical lenses.} Having established a similarity between these effects, he uses knowledge from geometrical optics to engineer a new account of energy flux in general relativity. The analogy thus established, and the thus created ability of thinking about Ricci curvature in particular as bringing about a positive focusing of light rays, would become a crucial tool for Penrose, among other things in his first singularity theorem of 1965, and in his argument that mass must be positive in 1993.

In the current paper, we analyze the details of Penrose's idea, how exactly his analogy between curved spacetime regions and optical lenses functions, and how far it goes. In particular, we examine the set of spacetimes for which the analogy holds, and use these results to investigate how the analogy might best be understood. Is it just a helpful but limited metaphor like the curved sheet of cloth, at best of heuristic value, or does it encode something deeper about the very concept of spacetime curvature?\footnote{In a later paper, we will investigate the roles the analogy played in Penrose's construction of the first 1965 singularity theorem in \citet{Penrose:1965d}, in his investigation of the breakdown of global hyperbolicity of plane wave spacetimes in \citet{Penrose1965:planewaves}, and in his proof of a positive mass theorem in \citet{penrose1993positive}. Some of these connections are already hinted at in Penrose's 1967 Battelle Lectures \citep{penrose-struc-st2}.} 

We will proceed in the following way. In Section \ref{SectionII}, we put Penrose's paper \citep{Penrose:1966a} on the analogy between curved spacetime regions and optical lenses in its historical context and introduce prior work that Penrose relies on. In Section \ref{SectionIII}, we give an overview of the aim and structure of his paper. In Section \ref{SectionIV} and Appendices A and B, we introduce the particular mathematical and physical concepts, formalisms, and tools that Penrose uses and that are relevant for understanding the details of his paper. Then, in Section \ref{SectionV}, we proceed with a detailed analysis of the explicit and implicit assumptions Penrose employs in the construction of his analogy, discuss its strengths and limitations, unpack and extend the details of his proof of concept, and examine Penrose's application of the analogy to three different types of spacetimes. Finally, resting on all this, we show in Section \ref{SectionVI} that his analogy can be well matched with the philosophical account of analogy based on Norton's material theory of induction.

\section{Historical Preliminaries} \label{SectionII}

\subsection{Prehistory} \label{SubsectionIIA}

\noindent The history of Penrose's analogy starts with the idea of the geometry of curved surfaces, which in turn starts with Carl Friedrich Gauss. Gauss developed many of the mathematical tools, most importantly that of a variable metric tensor, to describe the geometry of curved $2$-dimensional surfaces. Gauss' student Bernhard Riemann extended this work to manifolds of arbitrary dimension, and defined what we today call the Riemann curvature tensor, generalized from Gauss' sectional curvature of $2$-surfaces.\footnote{See the new edition of Riemann's habilitation thesis \citep{RiemannWeyl:1923} edited by Hermann Weyl. For the development of the concept of a manifold from Riemann onwards see \citet{Scholz:1980}.} However, it is often forgotten that after Riemann the development of these mathematical objects was embedded in a more general theory of bilinear forms, that went away from geometric interpretations and focused more on their algebraic and invariance properties. In the 1880s and 1890s, Ricci and Levi-Civita developed what was long called ``Ricci calculus,'' and would today be called ``tensor calculus,'' emphasizing that  Gauss' and Riemann's metric tensors are just special cases of a more general bilinear form. They defined the Riemann curvature tensor via the first- and second-order derivatives of such a form. Thinking of all these objects in a purely geometric way was therefore thought of as old-fashioned at best at the time, and nongeneral in the worst cases. Thus, when Einstein and Grossmann embarked on developing a tensorial theory of gravity in 1913, they did not think of the metric tensor and the Riemann tensor as being, first and foremost, about the geometry of spacetime but more as tools to express a classical field theory.\footnote{See \citet{Reich:1992, reich1994entwicklung} and \citet{Lehmkuhl2014:geo, Lehmkuhl:forthOUP} for details.} Although Einstein himself never properly came around to thinking about general relativity in a predominantly geometrical way, many others in the emerging community of general relativists soon did, essentially affecting a counter-revolution, especially after Levi-Civita had introduced the concept of parallel transport in 1917, and after Weyl generalized this concept and introduced the definition of the Riemann tensor in terms of such parallel transport in the first edition of his famed book \textit{Raum-Zeit-Materie} (Space-Time-Matter) in 1918.

Einstein studied both the proofs of the first edition of \textit{Raum-Zeit-Materie} and the first article in Weyl's pursuit of a unified field theory of gravity and electromagnetism \citep{Weyl:1918b} in April of 1918.\footnote{For details on Einstein's and Weyl's correspondence on this see \citet{Lehmkuhl:forthOUP}.} The latter was indeed the very first attempt at a unified field theory, and was the physicist-directed part of a pair of papers. The mathematician-directed companion part \citep{Weyl:1918}, however, not only introduces Weyl geometry, the mathematical foundation of Weyl's unified field theory, it also contains his definition of what would later be called the Weyl tensor,\footnote{In \citet{Einstein:1921e}, Einstein explicitly uses the Weyl tensor for the first time and names it thus (also possibly for the first time). Einstein was skeptical of Weyl geometry and Weyl's unified field theory as such (see \citet{Lehmkuhl:forthOUP} for Einstein's seven arguments against Weyl), but in this paper argues for using the Weyl tensor in the context of a theory based entirely on conformal geometry.} and shows how both the Ricci tensor and the Weyl tensor can be obtained from the Riemann tensor.\footnote{Weyl made these developments in the context of Weyl geometry, but the definition of the Weyl tensor is proposed in connection with what he calls \textit{directional curvature} (\textit{Richtungskr\"ummung}), i.e., the part of the total curvature tensor of Weyl geometry that coincides with the Riemann tensor of pseudo-Riemannian geometry.} Just a month before receiving the proof of \citet{Weyl:1918b}, Einstein had submitted \citet{Einstein:1918}, in which he first defines the term ``Mach's principle'' as the demand that the energy-momentum tensor ``uniquely determines'' the ``state of space described by the fundamental tensor,'' namely the metric tensor. It might seem puzzling that Einstein expected the energy-momentum tensor to determine the metric uniquely given that only the Einstein tensor, which is a combination of the Ricci tensor, the scalar curvature, and the metric, is determined by the Einstein field equations, whereas the part of the Riemann tensor that corresponds to the Weyl tensor is left constrained but not determined by the energy-momentum tensor.\footnote{See Equation (\ref{divergence}) in Section \ref{SubsubsectionIVB}.} But Einstein saw the metric as the gravitational potential, the connection as representative of the gravitational field, and the Einstein tensor as the rate of change of the gravitational field; he did not associate the rest of the Riemann tensor with gravity.\footnote{See \citet{Rennjanssen:2007}, \citet{RennSauer2006}, and \citet{norton2007einstein} for details.} And he saw a pseudo-tensor composed of first-order derivatives of the metric as the representative of gravitational energy \citep[cf.][]{Einstein:1916e}. 

We believe that it has not been quite appreciated how much the way of interpreting general relativity that became prominent from the 1950s onward---and today is so prominent that it is barely seen as an interpretation anymore but instead as a fact---differs from Einstein's interpretation. In particular, in contrast to Einstein's view, the nonvanishing of the Riemann tensor, i.e., the \textit{total} curvature tensor, is now often taken as the necessary and sufficient condition for a gravitational field to be present. J.\ L.~Synge was arguably the trailblazer of this new focus; in his influential 1960 textbook he wrote:\footnote{See \citet[pp.\ VIII--IX]{synge1960relativity}.}

\begin{quote}
If we accept the idea that spacetime is a Riemannian four-space (and if we are relativists we must), then surely our task is to get the feel of it (...). And the first thing we have to get the feel of is the Riemann tensor, for it \textit{is} the gravitational field---if it vanishes, and only then, there is no field. (...)
In Einstein's theory, either there is a gravitational field or there is none, according as the Riemann tensor does not or does vanish.
\end{quote}

\noindent As we shall see in the following subsection, Penrose was one of the proponents of this new way of seeing general relativity.

\subsection{The Historical Context of Penrose's Paper} \label{SubsectionIIB}

\noindent Roger Penrose decided to work on general relativity in early 1958. Until then, he had had a career in pure mathematics, having just finished his PhD thesis on ``Tensor Methods in Algebraic Geometry'' at St.\ John's College Cambridge in 1957; a thesis not connected to general relativity.\footnote{Penrose never published his thesis, but it is now available as Document 6 in \citet{penrosecollectedworksvolume1}.} However, throughout his time as a PhD student at Cambridge, he was in close contact with the cosmologist Dennis Sciama, whom he credits as having ``had the greatest influence on the development of my research during this period [as a PhD student], and for many years later.''\footnote{See Penrose's short autobiography in Volume 1 of his Collected Works, most likely written in 2010, and the immediately preceding preface. There, Penrose also notes the influence that Fred Hoyle's BBC radio talks on cosmology in 1951, Hermann Bondi's lectures on general relativity, and Paul Dirac's lectures on quantum mechanics had on him during his time as a PhD student, as well as interactions with Felix Pirani, who finished his second PhD thesis at Cambridge under Hermann Bondi in 1956.}  

It was Sciama who encouraged Penrose to attend a lecture by David Finkelstein at King's College London in early 1958. Finkelstein's talk was based on his paper \citep{Finkelstein:1958}, in which he, inter alia, argued that in the Schwarzschild solution of the vacuum Einstein field equations the alleged singularity at $r = 2m$ was not a ``real'' singularity but could be transformed away by changing the coordinate system.\footnote{Here, $r$ is the Schwarzschild radial coordinate and $m$ is standardly interpreted as the mass of the spherically symmetric body whose exterior gravitational field the Schwarzschild solution represents. For more on the history of the early interpretation of the Schwarzschild solution see \citet{eisenstaedt:1989,eisenstaedt1993lemaitre,eisenstaedt2006curious}.} Many years later, Penrose stated that it was this talk by Finkelstein that planted the seed of the question that would eventually lead to his first singularity theorem of 1965 \citep[see][]{Penrose:1965d}, and that it was this event that convinced him to change his research focus to general relativity. In these same interviews, Penrose also stated that he felt that he needed to approach general relativity in a somewhat ``quirky'' way that was not well-trodden by others; and having been impressed by Dirac's lectures on spinors in quantum mechanics, he set himself the task of reformulating general relativity in spinorial form.\footnote{For all this see Thorne's interviews with Penrose recorded in 1983 (Thorne uses them in \citet{Thorne:1995a}; the complete interviews are deposited at the Caltech Archives), Lightman's interview with Penrose in 1989 \citep{penrose1989AIPinterview}, and Stamp's and Lehmkuhl's interviews with Penrose in 2019.}

The first conference on general relativity and gravitation that Penrose took part in, but a year after his starting to work on general relativity, took place in Royaumont, France, from 21 June to 27 June 1959. It was the third conference in a newly introduced series of conferences specifically on general relativity and gravitation that had started in Berne, Switzerland, in 1955 and was followed by Chapel Hill, North Carolina, in 1957. The third conference of this series, that would be the seed from which the International Society of General Relativity and Gravitation grew,\footnote{The society was formally founded in 1971, and grew from the International Committee on General Relativity and Gravitation that was founded in 1957. For the history of these conferences and how they grew from the Berne conference see \citet{kiefer2020space} and \citet{Blumetal:2015,blum2016renaissance,blum2018gravitational}.} was hosted by M.\ A.~Lichnerowicz and M.\ A.~Tonnelat, and as the two predecessor conferences, the Royaumont conference brought together everybody who was anybody in general relativity and gravitational research. This list would not normally have had included Penrose, who had only recently graduated on a topic in algebraic geometry, and did not yet have any publications in the field of general relativity. However, Dennis Sciama sacrificed half of his slot to make it possible for Penrose to give a talk,\footnote{See Penrose's editorial note to Document 14 of \citet{penrosecollectedworksvolume1}; this paper is the written version of the talk he gave in Royaumont. The conference proceedings \citep{lichnerowicz-tonnelat-theors-rel-grav} only came out in 1962, and thus substantially after Penrose's later written ``A Spinor Approach to General Relativity'' \citep{Penrose:1960a}, which is a much more detailed version of the same paper.} and so Penrose presented the first fruits of his spinor reformulation of general relativity, including his spinorial representations of the Ricci and Weyl curvature tensors, and a spinorial version of the classification of vacuum spacetimes that had recently been pioneered by \citet{Petrov:1954} and used ingeniously by \citet{Pirani:1957} for a new account of what it is to be a  gravitational wave and what it means for such a wave to possess energy.\footnote{We will investigate the development and interpretation of the Petrov--Pirani--Penrose classification in a separate paper.} 

Both Petrov and Pirani were present at the Royaumont conference, as was J.\ L.~Synge, who likewise addressed the question of how gravitational energy ought to be defined. It was likely also the first time that Penrose met J\"urgen Ehlers and Rainer (Ray) Sachs, whose work would be a major launchpad for the paper with whose analysis we will be concerned with. Both Ehlers and Sachs were at the time members of Pascual Jordan's general relativity group at the University of Hamburg. Ehlers was 30 years old at the time of the conference, two years older than Penrose, and had finished his PhD with Jordan two years prior. In his PhD thesis, Ehlers had pioneered new classes of exact solutions to the Einstein field equations, new methods for finding solutions, as well as general relativistic hydrodynamics \citep{Ehlers:1958a}. Sachs was three years younger than Ehlers, had just finished his PhD thesis with Peter Bergmann at Syracuse University and must have moved to Hamburg to join Jordan's group mere months before the Royaumont conference took place.\footnote{This assumption is based on the fact that Sachs and Bergmann published a paper in Physical Review \citep{sachs1958structure} that was received on 12 June 1958, and the affiliation of both of them is named as Syracuse University; yet in the Proceedings of the Royaumont conference \citep{lichnerowicz-tonnelat-theors-rel-grav}, which took place in June 1959, Sachs' affiliation is given as Hamburg University.} Within a year, he published his first single-authored paper, which, he wrote, was inspired by the ``beautiful analogy pointed out by Pirani between electromagnetic and gravitational null fields'' \citep[][p.\ 465]{Sachs:1960a}.

Penrose's starting point for everything that follows in his paper on the gravito-optical analogy are what he calls ``Robinson's and Sachs' optical scalars,'' which constitute the first ingredient, arguably the all-important seed, from which Penrose would grow his far-reaching analogy between gravitational and optical systems. Penrose refers to a paper by \citet{Robinson:1961a}, one by \citet*{JordanEhlersSachs}, and one by \citet{Sachs:1961a} alone. Robinson's paper was submitted first (on 9 September 1960), but in it he is open about using results from Sachs on the optical scalars, though there was as of yet no paper to be cited. This changed with the paper by Jordan, Ehlers, and Sachs which first introduced the optical scalars systematically.\footnote{This paper by Jordan, Ehlers, and Sachs is the second in a series of papers published in the proceedings of the Mainz academy. The series of papers soon became known as ``the Hamburg bible,'' and unfolded an influence well beyond what had become normal for papers written in German rather than in English.} It was submitted 7 weeks after Robinson's paper, and introduced a variety of new results and methods on exact solutions to the Einstein field equations. Sachs submitted his own paper focusing entirely on the use of---and theorems concerning---the optical scalars only four weeks after the joint paper with Jordan and Ehlers, after he had moved on to Hermann Bondi's group at King's College London. Thus, even though the optical scalars were first mentioned in Robinson's paper (with reference to Sachs) and first fully introduced in the paper by Jordan, Ehlers, and Sachs, it seems plausible that the fact that they soon became known as ``Sachs' optical scalars'' is justified.\footnote{In a cosmological context, \citet{raychaudhuri1955relativistic} explicitly defines only the rotation for a congruence of timelike geodesics. To our knowledge, Sachs was the first to derive all three optical scalars and their evolution equations for geodesic null congruences.}

Having somewhat settled the question of priority, what \textit{are} Sachs' optical scalars, and how did he interpret them? In the first focused work on this, \citep{Sachs:1961a}, Sachs introduced the idea of exploring and indeed defining the properties of a given spacetime through congruences of geodesic null curves, i.e., families of curves in spacetime that can be interpreted as bundles of light rays. He then introduced the optical scalars as the \textit{convergence} $\theta$, the \textit{rotation} $\omega$, and the \textit{shear} $\sigma$ of said null congruence.\footnote{In \citet*{JordanEhlersSachs}, they are introduced as \textit{Divergenz}, \textit{Drill}, and \textit{Verzerrung}, respectively. Accordingly, the more appropriate term for the optical scalar $\omega$ would be \textit{twist}. However, we here stick to Penrose's nomenclature.} The precise definitions and geometrical meanings are presented in our Appendix B; here we will just give an intuitive picture of the optical scalars as introduced by Sachs that may well have fueled Penrose's imagination.\footnote{The following is a slightly embellished version of what \citet[p.\ 317]{Sachs:1961a} calls ``the shadow experiment.''}

Imagine sitting in a spherical spaceship that effectively moves, as all material bodies do, on a timelike curve of that spacetime. You are eager to determine the curvature properties of the spacetime you are moving through, but you do not know how. Then you enter Earth's solar system, passing between a space station with a flat surface on your left, and the Sun on your right. The rays of the Sun will hit your spherical spaceship, and throw a shadow on the flat surface of the space station. Your eyes widen because you realize that the \textit{shape} of that shadow allows you to determine in what way the spacetime region through which the light rays had moved were curved. If there was \textit{no} curvature in the intervening region, then the shadow of your spaceship will just be a circular disk. Sachs' optical scalars tell you that depending on the precise curving of the intervening spacetime region, the bundle of light rays that hit your spaceship, and thus the resulting shadow, can be \textit{contracted or expanded} (the amount of which is determined by $\theta$), it can be \textit{twisted} (determined by $\omega$), and it can be \textit{sheared} (determined by $\sigma$), i.e., be deformed from a circular to an elliptic disk. Accordingly, these light rays act as a kind of waywiser for the curvature distribution in the intervening spacetime region, and the shadow on the flat surface of the space station is your measurement output.\footnote{For the precise way in which the optical scalars are related to the Riemann tensor see the Sachs equations in Appendix~B.} 

We do not know if Penrose had heard the first thoughts on all this at the Royaumont conference; but it is plausible given that a series of detailed papers introducing and using Sachs' optical scalars already came out a year later. Either way, it is clear that Penrose paid apt attention to these publications, now from his new environment in the United States. For just a few weeks after the Royaumont conference, Penrose had moved to Princeton as a NATO Research Fellow, to join the burgeoning general relativity research group of John A.\ Wheeler.\footnote{See \citet[Section 5]{Rickles:2020a} for the role of NATO funding for gravitational physics in these years.} As he notes in the first footnote of the gravito-optical paper, this is where he started to work on the paper in 1960. 

It was some time during his two years as a NATO Research Fellow that Penrose first started working with Ezra Theodore (Ted) Newman---a collaboration that would span more than half a century. Both Penrose and Newman had given separate spinor reformulations of general relativity, but whereas Penrose was a geometrical thinker, Newman thought algebraically, and they found that these differences in approach complemented each other extremely well. They teamed up and created what soon became known as the Newman--Penrose formalism of general relativity, which combined many of the advantages of their originally separate approaches. Next to Sachs' optical scalars, the Newman--Penrose formalism was the second main ingredient of Penrose's gravito-optical paper, for it provided the language and methods of proof in which Penrose would spell out his arguments about the ways in which curved spacetime regions and optical lenses were analogous to each other.\footnote{See \citet{NewmanPenrose1962} for the original Newman--Penrose paper, and our Appendix A for all the elements of the formalism that are crucial for the arguments in Penrose's gravito-optical paper. Note also that at the end of the Newman--Penrose paper, there is a restatement of Sachs' peeling theorem in the language of the Newman--Penrose formalism, preparing later work on the energy of gravitational waves.}

Penrose returned from the United States in 1961, and became a Research Fellow at King's College London, where Hermann Bondi had started a general relativity group focused on gravitational wave research in 1955, directly inspired by the aftermath of the 1955 Berne conference. In London, Penrose's contact especially with Pirani and Sachs intensified, and it stands to reason that this had an influence on the further maturing of his paper on the gravito-optical analogy. Indeed, it would take until 1966 for the paper to come out in \citet{hoffmann1966perspectives}, a Festschrift for V\'{a}clav Hlavat\'{y} edited by Banesh Hoffmann, so that the later parts of the paper cite many sources that appeared in the meantime, including Penrose's first singularity theorem from 1965. Thus, though a major part of the paper was likely written in the early 1960s, Penrose did take advantage of including up to date references shortly before publication.

\section{Aim and Structure of Penrose's Paper} \label{SectionIII}

The aim and structure of Penrose's paper are as follows. In his introductory \textbf{Section 1, ``Non-Locality of Energy,''} he first states the main mission of the paper: to clarify ``that somewhat elusive concept---gravitational energy'' by drawing on considerations from elementary lens optics. The key idea that Penrose will exploit is that there is a crucial analogy between how light is influenced by passing through \textit{curved spacetime regions} (gravitational fields) on the one hand, and \textit{curved glass} (optical lenses) on the other hand. In particular, he will argue that two types of spacetime curvature present in general relativity---Ricci curvature and Weyl curvature---can act like two types of elementary lenses in geometrical optics. That is, he will demonstrate that, in the right circumstances, spacetime regions with nonvanishing Ricci and Weyl curvatures \textit{focus} light in just the same way as anastigmatic and astigmatic lenses, respectively, and that the amount of focusing can be used to measure the total energy-momentum flux associated with that curved spacetime region. 

In order to achieve this, Penrose starts, in his \textbf{Section 2, ``Optical Scalars,''} by picking out one particular null geodesic $n$ that belongs to a specifically arranged congruence of null geodesics in spacetime. Then, he defines Sachs' optical scalars $\theta$, $\omega$, and $\sigma$, which respectively measure the convergence, rotation, and shear of the null congruence as one traverses spacetime along $n$, depending on how curved the respective spacetime regions are.\footnote{More precisely, he combines the convergence $\theta$ and the rotation $\omega$ into one complex number $\rho$. None of his arguments depend on this move, however, and we have found it convenient in our reconstruction to stick to the original three optical scalars.} In more detail, to trace \textit{how} the optical scalars change as one traverses spacetime along the null geodesic $n$, he first introduces a propagation derivative $D$ that allows one to determine the change of a quantity along $n$, and then uses it to specify the Sachs equations, which describe how all three optical scalars change along $n$.\footnote{See Equations (\ref{ChangeOS1}), (\ref{ChangeOS2}), and (\ref{ChangeOS3}) in Appendix B.} Through the Sachs equations, he argues that the rate of change of the convergence $\theta$ and the rotation $\omega$ along $n$ is dominated by Ricci curvature, while the rate of change of the shear $\sigma$ along $n$ is dominated by Weyl curvature. Of course, only Ricci curvature is directly related to the energy-momentum distribution of matter, radiation, and nongravitational fields, $\boldsymbol{T}$,\footnote{However, we shall show in Section \ref{SubsubsectionIVB}, especially in and after Equation (\ref{divergence}), that Weyl curvature is, though not uniquely determined, influenced and constrained by $\boldsymbol{T}$ as well.} and Penrose uses this fact to argue that imposing what amounts to what would soon be called the null energy condition on $\boldsymbol{T}$ implies that the Ricci scalar $\Phi$ is necessarily nonnegative, while there is no such inequality for the Weyl scalar $\Psi$.\footnote{Penrose had introduced the spinor analogues of the familiar Ricci and Weyl tensors; see Appendix A for details. Here and in the following, when referring to ``the Ricci scalar'' $\Phi$ and ``the Weyl scalar'' $\Psi$, we mean the two scalars $\Phi_{0 0}$ and $\Psi_0$, as defined in Appendix B, that Penrose obtains from the Ricci spinor and the Weyl spinor, respectively. The Ricci scalar in \textit{this} sense needs to be sharply distinguished from what is also often called ``the Ricci scalar,'' namely the full trace of the Riemann tensor. We call the latter ``scalar curvature'' instead to avoid confusion.} 

\label{Psec3Intro} Subsequently, in \textbf{Section 3, ``Lenses,''} Penrose constructs a curved spacetime region that most clearly corresponds to an optical lens. As optical lenses are curved pieces of glass,\footnote{Of course, strictly speaking, optical lenses can be made of any refractive material, with plastic having recently superseded glass as the material that most bespectacled people wear; and indeed, Penrose never speaks of any particular material. Still, for nostalgic reasons we will often refer to optical lenses as curved pieces of glass in the following.} their effects on light are such that the light is unaffected until it enters the curved glass, and again unaffected once it leaves the curved glass on the other side. Analogously, Penrose assumes that the curved spacetime region he seeks is flat before and after the light ray enters the region that is curved in a particular way. He models this circumstance by setting both the Ricci scalar and the Weyl scalar proportional to Dirac delta distributions. This then allows him to argue that nonvanishing Ricci curvature in the spike region accompanied by vanishing Weyl curvature corresponds to one specific type of optical lens well-known from classical optics, whereas nonvanishing Weyl curvature accompanied by vanishing Ricci curvature corresponds to another such type of lens. In the first case, a purely Ricci-curved spike, he argues, the spacetime region in question has an effect on light that corresponds to that of an anastigmatic lens, i.e., it is like a \textit{magnifying lens} that \textit{focuses} the light rays passing through, makes them approach each other towards a single focal point.\footnote{The properties of both of anastigmatic and astigmatic lenses will be defined and explained with more precision in Section \ref{SubsectionIVB}.} Penrose then notes that the defining property of the second type of lens, an \textit{astigmatic lens}, is that in one plane the positive focusing of light rays having passed through it is exactly as great as the negative focusing in the perpendicular plane (p.\ 263).\footnote{In geometrical optics, the defining property of an astigmatic lens is, in short, the formation of \textit{two focal lines} behind the lens, the primary and secondary line images in FIG.\ 1 in Section \ref{SubsectionIVB}.} This, he says, is exactly what a purely Weyl-curved spike region of spacetime does. Thus, Penrose argues, while a purely Ricci-curved spike focuses light like a perfect magnifying glass into a single point, a purely Weyl-curved spike turns a light beam whose cross section is initially a circle into an ellipse and ultimately into a line. It is, thus, like a poorly-made magnifying glass that yields a blurry picture.\footnote{In Section \ref{SubsectionVC}, we shall argue that the analogy between Weyl-curved spacetime regions and astigmatic lenses is less perfect than between Ricci-curved spacetime regions and anastigmatic lenses. However, the consequence that is crucial for Penrose, namely that both astigmatic lenses and Weyl-curved spacetime regions turn a circular light beam into an elliptical one, remains true.} 

Penrose then observes that given all this, Ricci curvature and Weyl curvature ``are, in a sense ``orthogonal'' to each other'' (p.\ 264); but he also notes that thus far he had only been concerned  with ``first-order'' effects of \textit{local} focusing, and that a complete picture would demand doing justice to the nonlinearity of the Sachs equations. One might have expected that this is where the analogy breaks down, given that, as just described, the effect that a single lens has on light is a rather local affair. But Penrose now turns around and looks for a feature in geometrical optics that corresponds to the nonlinearity of the Sachs equations. He finds it in the fact that the total focusing power of \textit{two} anastigmatic lenses is only additive---and hence in a sense linear---if the two lenses are placed directly against each other. In contrast, their total focusing power becomes more and more nonlinear if they are placed at an increased distance from each other. Aiming to learn more about the properties of Ricci- and Weyl-curved spacetime regions by making further use of the analogy, Penrose thus explores the properties of elementary lenses in more depth. 

A major tool used later in his further construction of curved spacetime regions corresponding to optical lenses originates from what he calls ``a classical theorem of elementary optics'' (p.\ 265): that a \textit{system of} two thin, anastigmatic lenses is equivalent in the effects it has on light passing through it to a \textit{single} thick, anastigmatic lens. More surprisingly, though, and of even more use in his later transfer to curved spacetime regions, he finds that if two \textit{astigmatic} lenses are arranged in the right way, then the resulting focusing power is also as if we had a single thick, \textit{anastigmatic} lens. The analogous effect to be used later in the paper is then that in some cases Weyl-curved spacetime regions can be modeled by Ricci-curved spacetime regions. 

Having thus explored some of the properties of systems containing two thin (and convex), either anastigmatic or astigmatic lenses, Penrose makes a crucial step at the beginning of his \textbf{Section 4}, entitled \textbf{``Energy Flux as Focusing Power?''} He uses the effective additivity of the \textit{focusing powers} of two such anastigmatic lenses in cases where the distances involved are small compared to the focal lengths involved and that locally the net focusing powers of particularly arranged systems of two thin, astigmatic lenses is zero to argue that in the curved spacetime context something similar happens: for distances that are small compared to the radii of curvature involved in the Riemann tensor of the respective spacetime, the total \textit{energy-momentum flux} is effectively additive and the contribution of Weyl curvature to the total energy-momentum flux is essentially zero.\footnote{The old puzzle of \textit{why and in what sense precisely} gravitational energy, in contrast to the energy of matter, is nonlocal, can thus be linked to the properties of the Weyl tensor in spacetimes where the radii of curvature are \textit{not} small compared to the distances under consideration.} More precisely put, the analogical correspondence is between the focusing powers of optical lenses on the one hand, and the focusing powers of curved spacetime regions on the other, resulting in the correspondence between the focusing powers of lenses and, through Ricci curvature being linked to the energy-momentum tensor via the Einstein field equations, energy-momentum flux in curved spacetime regions: the focusing power \textit{of} the Ricci curvature of a spacetime region is a measure of the local total energy-momentum flux \textit{in} that region. However, as Penrose immediately reminds us, in cases where the focusing occurs over larger distances ``nonlinear (or nonlocal) effects must be taken into account'' (p.\ 266).

Using the comparison to systems consisting of two anastigmatic lenses that are not arbitrarily close to each other,\footnote{This will be elaborated in more detail in Section \ref{SubsectionVA}.} Penrose then spends the rest of his Section 4 to put the analogy to work in analyzing gravitational systems. He sets himself the aim of determining the focusing powers of---and thus the energy-momentum fluxes in---two different gravitational systems: (approximately) plane gravitational waves and plane-polarized gravitational wave packets. Both are modeled via particularly arranged series of the Weyl-curved Dirac delta regions of spacetime that he had first introduced when discussing the spacetime-counterpart of a \textit{single} thin, convex, astigmatic lens. In both cases, he makes his life significantly easier by use of the above-mentioned result from classical optics (or rather of its gravitational counterpart): he replaces pairs of Weyl-curved Dirac delta patches with single Ricci-curved patches; the equivalent of replacing two thin, convex, astigmatic lenses, arranged just so, with a single thick, convex, anastigmatic lens. And it is the total focusing powers of (the series of) such single Ricci-curved patches that Penrose takes as measures for the total energy-momentum fluxes within said gravitational systems.

It is quite impressive how far the analogy between curved spacetime regions and optical lenses has carried Penrose until this point in the paper. However, at this point he himself cautions the reader: ``The preceding arguments all have been concerned with focusing when the effective lenses are in some sense weak. The connection between energy-momentum flux and focusing power is, perhaps, not so clear when the lenses are strong'' (p.\ 269). Hence, in his final section of the paper, \textbf{Section 5}, entitled \textbf{``Focusing Power of a General System,''} he makes a first step towards addressing this problem, even though he had already noted in Section 1, ``Nonlocality of Energy,'' that ``[w]hether or not this idea [of connecting energy-momentum flux to focusing power] can be made completely rigorous and physically acceptable will depend on future developments.'' Still, Penrose wants to explore how far the analogy can be pushed. Indeed, in order to make this push possible, he first solidifies and makes precise his earlier claim that a Ricci-curved Dirac delta region of spacetime behaves like an anastigmatic lens and a Weyl-curved Dirac delta region like an astigmatic lens. Then, he aims to take the result beyond its original domain of single lenses: ``The general idea is to find quantities which measure correctly the focusing power when the system consists of a single lens and to use these same quantities to define the focusing power of a general system'' (p.\ 269).\footnote{The ``general system'' he has in mind is a spacetime region that contains the ray $n$, and is finite in extent along it.} 

Thus, he begins by splitting the light beam that comprises the null geodesic $n$ into ``the incident beam,'' i.e., the beam before it hits the curved spacetime region, and ``the emergent beam,'' the beam after it hits the curved region. Then, he uses the Sachs equations to trace the evolution of the optical scalars of the beam in order to compare the optical scalars of the incident and the emergent beam. This serves to justify the original idea of Penrose's analogy: comparing the optical scalars of a light beam before and after the beam enters the curved spacetime regions provides justification for seeing the curved region as analogous to an optical lens. For the corresponding solution to the Sachs equations shows that a light beam entering the curved region is influenced very much like a light beam entering an optical lens. In more detail, when formulating the Sachs equations, Penrose states, as already noted, that the evolution of the convergence $\theta$ and the rotation $\omega$ of the beam would be dominated by the Ricci scalar $\Phi$, whereas the evolution of the shear $\sigma$ would be dominated by the Weyl scalar $\Psi$ in the spacetime region that the light ray traverses. Implicitly assuming there to be no rotation (which is a necessary condition for general relativistic light rays to behave like light rays as conceptualized in geometrical optics),\footnote{See Sections \ref{SubsectionIVB} and \ref{SubsectionVA} for details.} Penrose gives an explicit solution for the evolution of convergence $\theta$ and shear $\sigma$ as the light beam passes through the curved spacetime region. Then, he sets himself to calculate the focal points of the null geodesics that make up the beam, viz., the focal points corresponding to the solution of $\theta$ and $\sigma$ of the Sachs equations. These are the points where neighboring null geodesics cross and where either $\theta$ or $\sigma$ tend to infinity. Penrose takes his results as showing that both purely Ricci-curved Dirac delta regions of spacetime and purely Weyl-curved Dirac delta regions influence light \textit{exactly} like anastigmatic lenses and astigmatic lenses, respectively, linking the evolution of the convergence $\theta$ to anastigmatic focusing and the evolution of the shear $\sigma$ to astigmatic focusing as the light beam traverses the curved spacetime region. 

However, our reconstruction in Section \ref{SubsectionVC} proves that his solution to the Sachs equations actually shows that while it is true that purely Ricci-curved Dirac delta regions influence light \textit{exactly} like anastigmatic lenses in that a geodesic null congruence converges into exactly one focal point, the influence of purely Weyl-curved Dirac delta regions is only \textit{similar} to astigmatic lenses. The reason is the following. An actual astigmatic lens exhibits \textit{two} orthogonal focal lines in the domain of the emergent beam. Yet, we found that Penrose's solution to the Sachs equations also in this case produces exactly \textit{one} focal point behind the lens. But this focal point turns out to be a line segment instead of a point, and even if there is only one such focal line present, it is this difference that makes the light ray behave similarly to how it would behave if it were traversing an astigmatic lens: the picture blurs as a circular light beam is deformed into a focal line instead of a focal point.\footnote{Thus, Penrose's overall argument is not hindered by this wrinkle. But we will see in Section \ref{SectionVI} that it teaches us a lot about how analogies in general, and Penrose's analogy in particular, can work.} 

In any event, the main point of Penrose's Section 5 was to perform the analysis of Dirac delta-curved spacetime regions in analogy to optical lenses in such a way ``that it would be applicable to more general systems'' (p.\ 270). That is why he then expresses his solution in such a manner that no \textit{explicit} knowledge of \textit{where} in spacetime the light beam might hit upon the Dirac delta-curved region is presumed, allowing him to motivate a definition of total energy-momentum flux across the beam that depends only on the optical scalars. In a nutshell, it makes precise the idea that the focusing power of a general system along---and thus its total energy-momentum flux across---a null geodesic up to a certain point can be measured by the focusing power of just a single lens evaluated at that point. Penrose is very open about the fact that he has justified his definition of energy only by appeal to very specific gravitational systems, but he argues that now not having to rely on knowledge of the position of the lens anymore allows him to ``envisage also using [this definition of total energy-momentum flux across the beam] in more general situations than that of a single lens'' (p.\ 271).  
In the remainder of this final section, Penrose compares the advantages, disadvantages, and relations of his new proposal for how to measure energy-momentum flux in general relativity with different alternatives, like those of defining total energy-momentum flux as the modulus or indeed the real part of the quantity he had found, the possibility of defining it by appeal to asymptotic spacetime structure, the idea of defining energy-momentum flux as such via energy-momentum flux at infinity, or defining it via Bondi's news function. 

Doing justice to these comparisons will not be the topic of this article, although it shows that at the time the notion of energy in general relativity was a subject of intense development. For now, we only want to get to the bottom of the details of the analogy between gravitational and optical systems that Penrose observes, and the manifold ways in which he uses and builds on the analogy. Moreover, we want to understand \textit{in what sense} the two sets of physical systems are analogous, and \textit{how far} the analogy goes. In order to achieve this, we will now, in Section \ref{SectionIV}, carefully introduce first the different types of curvature that Penrose draws on, and then the different types of optical lenses that he aims to link to particular curved spacetime regions. In Section \ref{SectionV}, we then unravel how precisely Penrose fashions the spacetime regions that mirror the effects that certain optical lenses have on light. We will reconstruct and go a bit beyond his analogy between Ricci-curved spacetime regions and anastigmatic lenses on the one hand, and Weyl-curved spacetime regions and astigmatic lenses on the other. Our analysis will show that rather than just transferring ideas, methods and results from one domain to the other, Penrose goes back and forth between gravitational and optical systems. Drawing on the details of this, in Section \ref{SectionVI}, we will investigate how Penrose's analogy fits with the account of analogy Norton proposed in the context of his material theory of induction. In doing so, we shall see that Penrose's analogy not only fits well with Norton's account, but that using the latter allows us to understand Penrose's account at a deeper level than would be possible without it. At the same time, the case study allows us to use Norton's account in such a way that does justice to Penrose's iterative way of going back and forth between the gravitational and the optical domains, and to see him along the way establish more than one ``fact of analogy,'' and make more than one ``analogical inference.''  

But one step at a time. Let us turn to the precise definitions and properties of types of curvature first.

\section{Mathematical and Physical Preliminaries} \label{SectionIV}

\noindent To be in a position to properly discuss Penrose's gravito-optical analogy, we first recall the relevant aspects of the canonical spacetime curvature tensors in general relativity. To be more precise, we recall their mathematical definitions and properties, the Ricci decomposition that relates these curvature tensors (and which is the starting point of Penrose's analysis), constraints on the curvature tensors, and their geometric interpretations. In doing so, we shall, however, allow ourselves to go slightly beyond the required scope, in particular in spelling out the interpretational possibilities of the different curvature tensors and their constraints, as they are so central for Penrose's analysis. Finally, we give a short account of the geometrical optics framework and of anastigmatic as well as astigmatic lenses, the other side of Penrose's analogy.

\subsection{Spacetime Curvature Tensors} \label{SubsectionIVA}

\subsubsection{The Riemann Tensor and its Contractions} \label{SubsubsectionIVA}

\noindent We begin by defining the canonical spacetime curvature tensors used in general relativity, namely the Riemann tensor, the Ricci tensor, the scalar curvature, and the Weyl tensor.\footnote{There are of course many other curvature tensors, such as the Kretschmann scalar, the Bel--Robinson tensor, and the Schouten tensor. For the purposes of the present study, however, we focus on the four standard curvature tensors.} To this end, we let $(\mathfrak{M}, \boldsymbol{g})$ be a $4$-dimensional Lorentzian manifold and $\boldsymbol{\nabla} \hspace{-0.07cm}: \Gamma(T\mathfrak{M}) \times \Gamma(T\mathfrak{M}) \rightarrow \Gamma(T\mathfrak{M})$ the Levi-Civita connection, where $\Gamma(T\mathfrak{M})$ is the space of smooth vector fields on $\mathfrak{M}$. The Riemann tensor $\boldsymbol{R} \hspace{-0.05cm} : \Gamma(T\mathfrak{M}) \times \Gamma(T\mathfrak{M}) \times \Gamma(T\mathfrak{M}) \rightarrow \Gamma(T\mathfrak{M})$ is an antisymmetric tensor field of type $(1, 3)$ defined as 
\begin{equation*} 
\boldsymbol{R}(\boldsymbol{X},\boldsymbol{Y}) \boldsymbol{Z} := \boldsymbol{\nabla}_{\boldsymbol{X}} \boldsymbol{\nabla}_{\boldsymbol{Y}} \boldsymbol{Z} - \boldsymbol{\nabla}_{\boldsymbol{Y}} \boldsymbol{\nabla}_{\boldsymbol{X}} \boldsymbol{Z} - \boldsymbol{\nabla}_{[\boldsymbol{X}, \boldsymbol{Y}]} \boldsymbol{Z} \, . 
\end{equation*} 
In a local coordinate system $\boldsymbol{x}$ on $\mathfrak{M}$, the components of the Riemann tensor may be expressed in the form
\begin{equation} \label{Riemann}
R\indices{^{\mu}_{\nu \alpha \beta}} = \partial_{\alpha} \Gamma\indices{^{\mu}_{\beta \nu}} - \partial_{\beta} \Gamma\indices{^{\mu}_{\alpha \nu}} + \Gamma\indices{^{\mu}_{\alpha \lambda}} \Gamma\indices{^{\lambda}_{\beta \nu}} - \Gamma\indices{^{\mu}_{\beta \lambda}} \Gamma\indices{^{\lambda}_{\alpha \nu}} \, ,
\end{equation} 
in which 
\begin{equation*} 
\Gamma\indices{^{\mu}_{\nu \alpha}} = \tfrac{1}{2} \, g^{\mu \beta} \hspace{0.03cm} (\partial_{\alpha} g_{\beta \nu} + \partial_{\nu} g_{\beta \alpha} - \partial_{\beta} g_{\nu \alpha})
\end{equation*} 
are the usual Christoffel symbols of the second kind. In what follows, we continue to employ local coordinate representations and drop the term \textit{components} for reasons of clarity and simplicity. The Riemann tensor possesses the skew symmetries $R_{\mu \nu (\alpha \beta)} = 0 = R_{(\mu \nu) \alpha \beta}$ as well as the interchange symmetry $R_{\mu \nu \alpha \beta} = R_{\alpha \beta \mu \nu}$. It furthermore satisfies the first and second Bianchi identities 
\begin{equation} \label{Bianchi}
R_{\mu [\nu \alpha \beta]} = 0 \quad \textnormal{and} \quad \nabla_{[\lambda} R_{\mu \nu] \alpha \beta} = 0 \, ,
\end{equation}
respectively. The partial traces of the Riemann tensor are either trivial, i.e., $R\indices{^{\mu}_{\mu \alpha \beta}} = 0 = R\indices{^{\mu}_{\nu \alpha \mu}}$ as a result of the skew symmetries, or give rise to the symmetric Ricci tensor $R_{\mu \nu} = R\indices{^{\alpha}_{\mu \alpha \nu}}$, which is the only nontrivial partial trace of the Riemann tensor. The full trace of the Riemann tensor leads to the scalar curvature $R = g^{\mu \nu} R_{\mu \nu}$. Using these quantities, one can directly define the fully trace-free part of the Riemann tensor, the so-called Weyl tensor, by
\begin{equation} \label{Weyl} 
C_{\mu \nu \alpha \beta} = R_{\mu \nu \alpha \beta} - g_{\mu [ \alpha} \, R_{\beta ] \nu} - g_{\nu [\beta} \, R_{\alpha ] \mu} - \tfrac{1}{3} \, g_{\mu [ \beta} \, g_{\alpha] \nu} \, R \, .
\end{equation} 
This curvature tensor naturally satisfies the same symmetries as the Riemann tensor, the first Bianchi identity in Equation (\ref{Bianchi}), and is per definition fully trace-free (i.e., $C\indices{^{\mu}_{\nu \mu \beta}} = 0$ and $C\indices{^{\mu}_{\mu \alpha \beta}} = 0 = C\indices{^{\mu}_{\nu \alpha \mu}}$ trivially again due to the skew symmetries). Equation (\ref{Weyl}) may also be viewed as a decomposition of the Riemann tensor into its irreducible representations under the action of the orthogonal group, coined the Ricci decomposition, given by
\begin{equation} \label{RicciDecomposition} 
R_{\mu \nu \alpha \beta} = C_{\mu \nu \alpha \beta} + E_{\mu \nu \alpha \beta} + S_{\mu \nu \alpha \beta} \, ,
\end{equation} 
in which the tensor 
\begin{equation} \label{E} 
E_{\mu \nu \alpha \beta} := g_{\mu [ \alpha} \, R_{\beta ] \nu} + g_{\nu [\beta} \, R_{\alpha ] \mu}
\end{equation} 
comprises only Ricci tensor contributions and the tensor  
\begin{equation} \label{S}
S_{\mu \nu \alpha \beta} := \tfrac{1}{3} \, g_{\mu [ \beta} \, g_{\alpha] \nu} \, R
\end{equation} 
consists of only scalar curvature contributions.\footnote{This representation of the Ricci decomposition is equivalent to Penrose's formulation given in Equations (1) and (2) on pages 259 and 260 of \citet{Penrose:1966a}, respectively.} Similarly to the Weyl tensor, these tensors possess the same symmetries as the Riemann tensor and satisfy the first Bianchi identity. Furthermore, the components of this decomposition are orthogonal to each other in the sense that\footnote{As a consequence, the Kretschmann scalar takes the simple form $K = R_{\mu \nu \alpha \beta} R^{\mu \nu \alpha \beta} = C_{\mu \nu \alpha \beta} C^{\mu \nu \alpha \beta} + E_{\mu \nu \alpha \beta} E^{\mu \nu \alpha \beta} + S_{\mu \nu \alpha \beta} S^{\mu \nu \alpha \beta}$.} 
\begin{equation*} 
C_{\mu \nu \alpha \beta} E^{\mu \nu \alpha \beta} = C_{\mu \nu \alpha \beta} S^{\mu \nu \alpha \beta} = E_{\mu \nu \alpha \beta} S^{\mu \nu \alpha \beta} = 0 \, .
\end{equation*} 
For more detailed information on the Ricci decomposition see, e.g., \citet[Chapter 1]{Besse} and \citet{SingerThorpe}. We point out that the Riemann tensor can be decomposed in many useful ways other than the one given by the Ricci decomposition. A prominent example is the so-called Bel decomposition \citep{Bel, Matte}, where the Riemann tensor is decomposed into three tensors that exhibit properties similar to those of electric and magnetic fields. For the purpose of the present paper, however, only the Ricci decomposition is required.

\subsubsection{Constraints on the Ricci and Weyl Tensors} \label{SubsubsectionIVB}

\noindent Within the framework of classical general relativity, one finds both the Ricci tensor and the Weyl tensor to be constrained by the energy-momentum tensor $T_{\mu \nu}$. To be more precise, in this framework, the Ricci tensor can be viewed as being constrained by the Einstein field equations in trace-reversed form
\begin{equation} \label{Einstein}
R_{\mu \nu} = T_{\mu \nu} - \tfrac{1}{2} \, g_{\mu \nu} T \, ,
\end{equation} 
where $T = g^{\mu \nu} T_{\mu \nu}$ is the trace of the energy-momentum tensor.\footnote{This is actually the original form of the Einstein field equations that Einstein published in late November 1915 \citep{Einstein:1915i}. The addition of the trace of the energy-momentum tensor was the all-deciding improvement on the field equations of Einstein's first paper of November 1915 \citep{Einstein:1915f}, which had forced him to presuppose that all matter accounted for by $T_{\mu \nu}$ is electromagnetic in nature so that $T = 0$ in the second November paper \citep{Einstein:1915g}. With the field equations (\ref{Einstein}), Einstein could now claim that his new theory of gravity placed no constraints on the nature of matter, a point he made emphatically when contrasting his theory with Hilbert's approach.} To see how the Weyl tensor is constrained, we substitute Equation (\ref{Weyl}) into the second Bianchi identity in Equation (\ref{Bianchi}), yielding \citep[cf., e.g.,][Chapter 4.1]{HawkingEllis}
\begin{equation*}
\nabla^{\mu} C_{\mu \nu \alpha \beta} = \nabla_{[\alpha} R_{\beta] \nu} + \tfrac{1}{6} \, g_{\nu [\alpha} \nabla_{\beta]} R \, .
\end{equation*} 
Then, using Equation (\ref{Einstein}), this relation immediately results in 
\begin{equation} \label{divergence}
\nabla^{\mu} C_{\mu \nu \alpha \beta} = \nabla_{[\alpha} T_{\beta] \nu} + \tfrac{1}{3} \, g_{\nu [\alpha} \nabla_{\beta]} T \, .
\end{equation} 
Consequently, by choosing a specific energy-momentum tensor, this first-order partial differential equation constrains the allowed functional shapes of the Weyl tensor, whereas the Ricci tensor is fully determined, as Equation (\ref{Einstein}) relates it algebraically to the energy-momentum tensor. 

Together with the Ricci decomposition, the above constraints on the Ricci and Weyl tensors suggest the following hierarchical structure of the canonical spacetime curvature tensors within general relativity: The central mathematical object describing the full gravitational field is the Riemann tensor (\ref{Riemann}) that, by definition, satisfies the first and second Bianchi identities, which, however, do not constrain it in any way. Via the Ricci decomposition (\ref{RicciDecomposition}), this tensor can be split into the trace-free part (\ref{Weyl}) and a part with nonvanishing trace given by (\ref{E}) and (\ref{S}). These two parts of the Riemann tensor, which account for different degrees of freedom of the gravitational field, are not of arbitrary form, but for a given energy-momentum tensor are subject to the divergence constraint (\ref{divergence}) on the one hand, and the Einstein constraint (\ref{Einstein}) on the other, leading to a particularly constrained Riemann tensor. The aspect that the Einstein field equations may be seen only as a constraint on the Ricci tensor, rather than a separate evolution equation, can be directly motivated by the contracted second Bianchi identity 
\begin{equation*}
\nabla_{\mu} R\indices{^{\mu}_{\nu}} = \tfrac{1}{2} \, \nabla_{\nu} R \, , 
\end{equation*}
which is a result of pure Riemannian geometry that does not rely on any input from general relativity. More precisely, rewriting this identity in the form 
\begin{equation*}
\nabla_{\mu} \bigl(R\indices{^{\mu}_{\nu}} - \tfrac{1}{2} \, g^{\mu}_{\nu} R\bigr) = 0
\end{equation*}
shows that the Einstein tensor $R\indices{^{\mu}_{\nu}} - \tfrac{1}{2} \, g^{\mu}_{\nu} R$ is divergence-free for all values of $\nu$, and therefore conserved. Accordingly, this puts a constraint on the Ricci tensor that, in general relativity, is identified with the right hand side of Equation (\ref{Einstein}).

\subsubsection{Geometric Interpretation} \label{SubsubsectionIVC}

\noindent Next, we discuss an interpretation of the canonical spacetime curvature tensors that relates to the geometric interpretation of trace functions, and comment on some of the current standard interpretations of the Weyl tensor. Although most of these views are well-known, our presentation gives rise to a coherent picture of the canonical spacetime curvature tensors. For different geometrical or physical interpretations see, e.g., \citet{BaezBunn}, \citet{Loveridge}, and \citet[Lectures 8 and 9]{Feynman}.

Firstly, we develop the notions and the geometric understanding of trace functions. For this purpose, we let $\mathcal{V}$ be a finite-dimensional inner product space and $\mathcal{H} \hspace{-0.07cm} : \mathcal{V} \rightarrow \mathcal{V}$ a linear operator. We denote the space of all linear operators on $\mathcal{V}$ by $L(\mathcal{V})$. Moreover, we define the time evolution operator $U \hspace{-0.07cm} : \mathcal{V} \rightarrow \mathcal{V}$ generated by $\mathcal{H}$, which is a smooth function in time $t$, via
\begin{equation} \label{flow} 
\begin{cases}
\, \displaystyle \frac{\textnormal{d}}{\textnormal{d}t} U(t) = \mathcal{H} \, U(t) \\[0.25cm]  
\hspace{0.35cm} U_{|t = 0} = \1_{\mathcal{V}} \, .
\end{cases} 
\end{equation}
We then consider any measurable set $S \subseteq \mathcal{V}$ and let $S_t = U(t) S$. Applying Jacobi's formula together with Equation (\ref{flow}), and using the fact that the determinant can be interpreted as a relative volume expansion, we readily find that 
\begin{equation*}
\frac{\textnormal{d}}{\textnormal{d}t}\textnormal{Vol}(S_t) = \textnormal{Tr}(\mathcal{H}) \, \textnormal{Vol}(S_t) \, ,
\end{equation*}
where $\textnormal{Vol} \hspace{-0.07cm} : \mathcal{V} \rightarrow \mathbb{R}_{> 0}$ with $\textnormal{Vol}(S_t) = \textnormal{det}\bigl(U(t)\bigr)$ is the volume of an $\textnormal{dim}(\mathcal{V})$-parallelotope, and $\textnormal{Tr} \hspace{-0.07cm} : L(\mathcal{V}) \rightarrow \mathbb{R}$ with $\textnormal{Tr}(\mathcal{H}) = \sum _{k \in I} \langle \mathcal{H} \, \boldsymbol{e}_k, \boldsymbol{e}_k\rangle$ is the usual trace of the matrix representation of the linear operator $\mathcal{H} \in L(\mathcal{V})$ with respect to the inner product $\langle \, . \, , \, . \, \rangle$ of $\mathcal{V}$ for an arbitrary basis $(\boldsymbol{e}_k)_{k \in I}$. Accordingly, the trace of $\mathcal{\mathcal{H}}$ can be seen as a measure of the rate of change of the volume of the image of the set $S$ under $U$. For a finite-dimensional, multilinear inner product space $\otimes_{l = 1}^m \mathcal{V}_l$, $m \in \mathbb{N}$, and for all $\mathcal{H}_j \in L(\mathcal{V}_j)$ and $j \in \{1, ... \, , m\}$, we can generalize this trace to the so-called partial trace over $\mathcal{V}_j$
\begin{equation*}
\textnormal{Tr}_{\mathcal{V}_j} \hspace{-0.07cm} : L\bigl(\otimes_{l = 1}^m \mathcal{V}_l\bigr) \rightarrow L\bigl(\otimes_{l = 1, l \not= j}^m \mathcal{V}_l\bigr) \quad \textnormal{with} \quad \textnormal{Tr}_{\mathcal{V}_j} \bigl(\otimes_{l = 1}^m \mathcal{H}_l\bigr) = \textnormal{Tr}(\mathcal{H}_j) \, \otimes_{l = 1, l \not= j}^m \mathcal{H}_l
\end{equation*}
as well as to the full trace 
\begin{equation*}
\textnormal{Tr}_{\otimes \mathcal{V}_l} \hspace{-0.07cm} : L\bigl(\otimes_{l = 1}^m \mathcal{V}_l\bigr) \rightarrow \mathbb{R} \quad \textnormal{with} \quad \textnormal{Tr}_{\otimes \mathcal{V}_l}\bigl(\otimes_{l = 1}^m \mathcal{H}_l\bigr) = \Pi_{l = 1}^m \textnormal{Tr}(\mathcal{H}_l) \, .
\end{equation*}
These traces may be viewed as measuring, in a sense, $\mathcal{V}_j$-directional and total rates of change of the volume of the image of a set $S \subseteq \otimes_{l = 1}^m \mathcal{V}_l$, respectively, under a generalization of the time evolution operator $U$ defined in Equation (\ref{flow}).

Now, we are in a position to give geometric accounts of the canonical spacetime curvature tensors. We begin with the standard geometric interpretation of the Riemann tensor in terms of the concept of geodesic deviation \citep[see, e.g.,][Chapter 8.7.]{mtw}. To this end, we consider a geodesic congruence $\bigl(\gamma_n(\tau)\bigr)_{n \in \mathbb{R}}$ with points $\mathfrak{M} \ni x^{\mu} = x^{\mu}(\tau, n)$, where $\tau$ is an affine parameter along---and $n$ a label for---the geodesics. Furthermore, we let $u^{\mu} = \textnormal{d}x^{\mu}/\textnormal{d}\tau$ be a tangent vector field to the geodesic congruence, and $\xi^{\mu} = \textnormal{d}x^{\mu}/\textnormal{d}n$ an infinitesimal displacement vector for neighboring geodesics measuring their separation. Then, the equation of geodesic deviation, which describes the relative acceleration of neighboring geodesics, reads
\begin{equation*}
u^{\alpha} \nabla_{\alpha} \bigl(u^{\beta} \nabla_{\beta} \xi^{\mu}\bigr) = \bigl(\partial_{\alpha} \Gamma\indices{^{\mu}_{\beta \nu}} - \partial_{\beta} \Gamma\indices{^{\mu}_{\alpha \nu}} + \Gamma\indices{^{\mu}_{\alpha \lambda}} \Gamma\indices{^{\lambda}_{\beta \nu}} - \Gamma\indices{^{\mu}_{\beta \lambda}} \Gamma\indices{^{\lambda}_{\alpha \nu}}\bigr) \, u^{\nu} u^{\alpha} \xi^{\beta} \, .
\end{equation*} 
Defining the Riemann tensor $R\indices{^{\mu}_{\nu \alpha \beta}}$ as the expression in the parentheses on the right hand side of this equation, it can be viewed simply as a measure of the (tidal) gravitational forces producing the relative acceleration between neighboring geodesics. Within this setting, and using the above geometric interpretations of the partial and full trace functions, where $S$ is now a subset of the (indefinite) inner product space $\otimes_{l = 1}^3 \mathcal{V}_l = \Gamma^3(T\mathfrak{M})$, and $\otimes_{l = 1}^3 \mathcal{H}_l = R_{\mu \nu \alpha \beta}$ the multi-linear operator, we can regard the Ricci tensor and the scalar curvature, which are the nontrivial partial trace and the full trace of the Riemann tensor with respect to the metric $g_{\mu \nu}$, respectively, as measures of the rates of $\Gamma(T\mathfrak{M})$-directional and total changes of volumes of small geodesic parallelotopes propagated along the geodesic null congruence. The Weyl tensor on the other hand, as the fully trace-free part of the Riemann tensor, does not contain any information on either directional or total changes of volumes of small geodesic parallelotopes. 

Its current standard interpretation, however, is based on the role it plays for certain vacuum spacetimes \citep[see, e.g.,][Appendix VI 3]{Szekeres, ChoquetBruhat}. More precisely, since for vacuum spacetimes $T_{\mu \nu} = 0$ everywhere, it follows from the Einstein field equations (\ref{Einstein}) that both $R_{\mu \nu} = 0$ and $R = 0$. Hence, the Ricci decomposition (\ref{RicciDecomposition}) yields the simple relation $R_{\mu \nu \alpha \beta} = C_{\mu \nu \alpha \beta}$, which means that for vacuum spacetimes the Weyl tensor comprises all information about the spacetime curvature.\footnote{Here, one assumes the trivial conditions $E_{\mu \nu \alpha \beta} = 0 = S_{\mu \nu \alpha \beta}$. However, it can be easily shown that also the more general condition $E_{\mu \nu \alpha \beta} = - S_{\mu \nu \alpha \beta}$ is satisfied only by the vacuum solution.} Moreover, for vacuum spacetimes, the divergence constraint (\ref{divergence}) on the Weyl tensor reduces to $\nabla^{\mu} C_{\mu \nu \alpha \beta} = 0$, holding for all four indices of the Weyl tensor due to its symmetries. Combining all this, one may regard the Weyl tensor as that part of the Riemann tensor which accounts for the curvature generated either by gravitational radiation traveling through regions of spacetime containing neither matter nor nongravitational fields, with the just mentioned vacuum divergence constraint as the associated propagation equation, or by localized, static sources as in the Schwarzschild solution, where the divergence constraint now assumes the role of a flux equation for the Weyl tensor implying that it is sink- and source-free, similarly to the case of the divergence constraint for the magnetic field in electromagnetism.\footnote{There exist, however, some cases, such as exact vacuum solutions of the Einstein field equations with both azimuthal and axial symmetries, for which this interpretation fails \citep{Hofmann_Niedermann_Schneider}.} This particular view of the Weyl tensor is often expected to be also valid for arbitrary nonvacuum spacetimes comprising regions where $T_{\mu \nu} \not= 0$, which is, however, unfounded as the full divergence constraint (\ref{divergence}) implies that the allowed functional shapes of the Weyl tensor are directly influenced by the energy-momentum tensor, that is, different nonvanishing energy-momentum tensors give rise to different constraints on---and thus different admissible functional shapes for---the Weyl tensor than in the vacuum case. As information on matter and nongravitational fields is encoded into the Weyl tensor in this particular sense, a physical interpretation that relies only on vacuum contributions to the solutions seems therefore insufficient. We note in passing that since the Weyl tensor is conformally invariant, i.e., under any conformal mapping of the metric $g_{\mu \nu} \mapsto g'_{\mu \nu} = \Omega^2 g_{\mu \nu}$ with $\Omega \in C^2_0(\mathfrak{M}, \mathbb{R} \backslash \{0\})$ the transformed Weyl tensor is of the form $C'_{\mu \nu \alpha \beta} = C_{\mu \nu \alpha \beta}$, it can also be viewed as a measure of the deviation of an arbitrary spacetime with respect to a locally conformally flat spacetime, for which the Weyl tensor vanishes identically.

\subsection{Geometrical Optics and Optical Lenses} \label{SubsectionIVB}

\noindent Geometrical optics is a particular short-wavelength approximation of wave optics, in which the wavelengths of light are small compared to the sizes of the interaction structures (see, e.g., \citet[Chapter III]{Born-Wolf} and \citet[Chapter 3]{Sears}). Hence, light propagation may be described in terms of rays, i.e., curves that are perpendicular to the wavefronts and obey Fermat's principle, that is, the path taken by a light ray between two given points is the one for which the traversal time is invariant with respect to small variations of the path. This approximation entails, however, that optical effects such as diffraction and interference cannot be accounted for. Nonetheless, geometrical optics may be employed to properly describe various aspects of imaging systems such as lenses, which are devices inducing a convergence or divergence of light rays due to refraction. This includes the effect of optical aberration, where light rays emerging from any given point on the imaging object do not converge at a single point in the image plane after transmission through the imaging system, causing the formation of blurred and/or distorted images. 

\begin{figure}[t]%
\centering
\includegraphics[width=0.65\columnwidth]{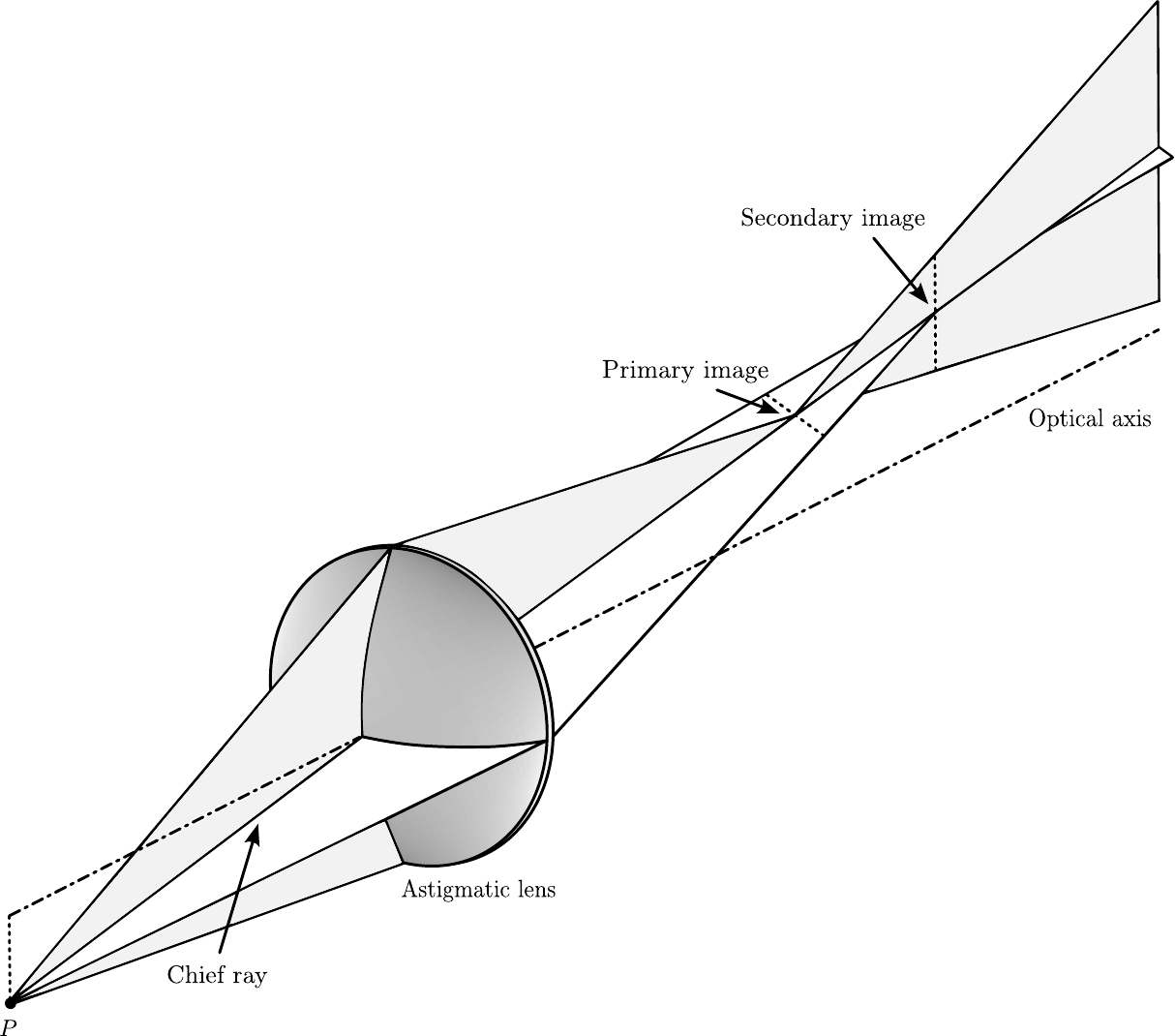}%
\caption[...]
{Schematic representation of an astigmatic lens and its effects. The cone of light rays, which is represented by two of its sections, emanating from an off-axis object point $P$ is refracted by the lens. After the refraction, all light rays intersect a horizontal line, the primary image, and subsequently a vertical line, the secondary image.}%
\label{lenses}%
\end{figure}

The particular optical imaging systems on which Penrose builds his study of the concept of energy in general relativity (cf.\ Section \ref{SubsectionVD})---and therefore to which he applies his gravito-optical analogy---are systems consisting of certain arrangements of two thin, convex, either anastigmatic or astigmatic lenses. In more detail, every thin, convex lens, for which the thickness is negligible compared to the curvature radius of the lens surface and which causes parallel-propagating light rays to converge, may exhibit the effects of optical aberration (which is of course also true for thick, convex lenses). The particular form of optical aberration considered by Penrose is the so-called astigmatic aberration,\footnote{For other forms of optical aberration such as spherical aberration, coma, or distortion see \citet[Chapter V]{Born-Wolf} or \citet[Chapter 5]{Sears}.} in which light rays that emerge from an object point not located on the optical axis and that lie either, e.g., in the meridional or the sagittal plane\footnote{The meridional plane is the plane defined by the object point and the optical axis, whereas the sagittal plane is perpendicular to the meridional plane and intersects the chief ray, that is, the ray connecting the object point and the center of the lens.} are refracted by the lens to different degrees. As a consequence, instead of producing one focused image point, both sets of light rays intersect the chief ray at different meridional or sagittal image points that extend as sharp, elongated line images oriented in the direction of the respective other plane, where the image formed in the meridional plane precedes the image formed in the sagittal plane. For light rays that are not contained in the meridional or sagittal planes, the image points coincide with the meridional and sagittal image points (for a schematic representation of the general setup see FIG.\ 1). Accordingly, for lenses exhibiting astigmatic aberration, the image is spread in the direction along the optical axis, where the best focus is found at the circle of least confusion located between the meridional or sagittal image points. Lenses that do not suffer from astigmatism, that is, lenses for which the sagittal and meridional image points are in alignment, are simply called anastigmatic. Now, placing two thin, convex, anastigmatic lenses with respective focusing powers $p_1$ and $p_2$ at a relative distance $w \geq 0$ to each other, the focusing power of the combined system becomes
\begin{equation} \label{focusingpower}
p = p_1 + p_2 - p_1 p_2 w \, . 
\end{equation}
Such a combined system effectively behaves like a single thick, convex, anastigmatic lens, which is characterized by its focal length $L = p^{- 1}$ as well as by the distance between its two principal planes,\footnote{The principal planes are hypothetical planes in a thick lens or a lens system that are perpendicular to the optical axis and contain all refraction loci. They are thus the planes of maximal focusing power.} here given by 
\begin{equation*}
k = - \frac{p_1 p_2}{p} w^2 \, .   
\end{equation*}
Lens systems of this kind serve as the fundamental building block in Penrose's effective construction of the more complex systems consisting of two thin, convex, astigmatic lenses (cf.\ Conditions ($\mathcal{O}$2)--($\mathcal{O}$4) in Section \ref{SubsectionVA}).

\section{Penrose's Gravito-optical Analogy} \label{SectionV}

\subsection{Outline and Assumptions} \label{SubsectionVA}

\noindent As discussed in detail in Section \ref{SectionIII}, Penrose's gravito-optical analogy first arises in one of his many studies of the concept of energy in general relativity, namely in \citet{Penrose:1966a}, where he presents a novel view on the topic suggesting that the resultant focusing power of spacetime curvature along certain geodesic null congruences is a suitable measure of the total energy-momentum flux, which also includes the energy-momentum flux of gravitation. To make this idea more precise, he considers a particular class of spacetimes with Ricci and Weyl tensors modeled by Dirac delta distributions, and uses the optical scalars associated with said geodesic null congruences as well as the Sachs equations expressed in the $2$-spinor representation of the Newman--Penrose formalism (cf.\ Appendices A and B) to draw a connection to the focusing power of certain elementary types of optical lenses within the geometrical optics framework. This allows him to show that the focusing effect of the (trace-free part of the) Ricci tensor is like that of a convex, anastigmatic lens, whereas the focusing effect of the Weyl tensor is like that of a convex, astigmatic lens. Although Penrose expects this behavior to hold for rather general spacetimes, he shows that it is definitely correct only in a certain limiting sense.

To analyze this in more detail, we first state Penrose's conditions on gravitational systems:
\begin{itemize}
\item[($\mathcal{G}$1)] The focusing power of spacetime curvature is measured only along geodesic null congruences [page 261, line 7].
\item[($\mathcal{G}$2)] The geodesic null congruences are devoid of any rotation, i.e., $\omega$ vanishes identically [page 263, lines 1--5].
\item[($\mathcal{G}$3)] The geodesic null congruences are nondispersive [page 263, lines 11--12].
\item[($\mathcal{G}$4)] The propagation equations for the convergence $\theta$ and the shear $\sigma$ along geodesic null congruences are the Sachs equations, in which the Ricci and Weyl scalars $\Phi$ and $\Psi$ are the dominant contributions [page 261, Equations (10) and (11); page 263, lines 14--15].
\item[($\mathcal{G}$5)] The Ricci scalar $\Phi$ is nonnegative, which amounts to imposing the null energy condition [page 262, lines 6--7].
\item[($\mathcal{G}$6)] The Ricci and Weyl scalars are modeled as 
\begin{equation} \label{RWD} 
\Phi = \Phi_{\delta} \hspace{0.02cm} \delta(r - r_0) \quad \textnormal{and} \quad \Psi = \Psi_{\delta} \hspace{0.02cm} \delta(r - r_0) \, ,
\end{equation}
respectively, where $\Phi_{\delta} \in \mathbb{R}_{> 0}$ and $\Psi_{\delta} \in \mathbb{C}$ are constants and $\delta(\, . \,)$ is the usual Dirac delta distribution with $r \in \mathbb{R}$ being an affine parameter along the geodesic null congruence [page 263, Equation (18)].
\end{itemize}
He notes that as a consequence of Condition ($\mathcal{G}$2), the geodesic null congruences are hypersurface-orthogonal, making them generators for systems of null hypersurfaces that can be seen as representing wave front sets for zero rest mass radiation. And together with Condition ($\mathcal{G}$3), these wave front sets behave like beams of rays, which can be described with a theory similar to geometrical optics. Furthermore, Conditions ($\mathcal{G}$5) and ($\mathcal{G}$6) impose restrictions allowing only for gravitational systems that resemble thin, convex lenses embedded in flat spacetimes. We also point out that Penrose seems to see Condition ($\mathcal{G}$4) more as a fact rather than an extra constraint on the nature of gravitational systems [page 263, beginning of Section 3]. This view is, however, unfounded, as the assumption that the dominant contributions in the Sachs equations are the Ricci and Weyl scalars $\Phi$ and $\Psi$ does not hold for arbitrary spacetimes, even if one just considers rotation-free geodesic null congruences. Accordingly, it is true only for a limited set of spacetimes, rendering it an imposed condition rather than a pure fact.

In addition to these conditions on gravitational systems, Penrose imposes the following conditions on optical lens systems:
\begin{itemize}
\item[($\mathcal{O}$1)] Over distances that are small compared to the focal length, the focusing powers of the constituents of a system consisting of two thin, convex, anastigmatic lenses are \textit{effectively} additive, i.e., locally the mixed term in the total focusing power (\ref{focusingpower}) can be neglected [page 265, lines 42--43]. For focusing over larger distances, the mixed term has to be taken into account [page 266, lines 6--10].
\item[($\mathcal{O}$2)] Systems consisting of two thin, convex, astigmatic lenses are restricted to those having identical and parallel-oriented constituents that are rotated through a right angle relative to each other in the plane of the lenses [page 265, lines 17--21].
\item[($\mathcal{O}$3)] Systems of two thin, convex, astigmatic lenses are described effectively by only the behavior of their principal planes [page 265, lines 21--23]. 
\item[($\mathcal{O}$4)] As the orthogonally-projected positions of the principal planes of systems of two thin, convex, astigmatic lenses on the meridional plane vary over a certain distance, a sufficiently large scale $\varepsilon \gg k/p$ is considered [page 265, lines 30--37].
\end{itemize}
As Penrose remarks, placing the lenses in Condition ($\mathcal{O}$2) directly against each other yields a zero total focusing power, whereas placing them so that they have a relative distance $w > 0$, one may use Equation (\ref{focusingpower}) for the total focusing power in each principal plane. Moreover, due to Condition ($\mathcal{O}$4), the principal planes appear to be flat. Thus, with Conditions ($\mathcal{O}$2)--($\mathcal{O}$4), the focusing effect of the more complicated case of a system consisting of two thin, convex, astigmatic lenses can be treated in essentially the same way as the focusing effect of an ordinary single thick, convex, anastigmatic lens. 

Taken together, the above conditions on optical lens systems translate into the following condition for their gravitational counterparts:
\begin{itemize}
\item[($\mathcal{G}$7)] Over distances much smaller than the radii of curvature involved in the Riemann tensor, the focusing powers exerted by the curvatures of two gravitational lenses as specified by Conditions ($\mathcal{G}$4)--($\mathcal{G}$6) along geodesic null congruences can be treated as \textit{effectively} additive, with the contribution of the Weyl tensor being essentially zero [page 265, lines 45--46; page 266, lines 1--2]. For larger distances, the nonlinearity of the Sachs equations has to be taken into account [page 266, second paragraph].
\end{itemize}
Thus, the total focusing power of said system over distances much smaller than the radii of curvature involved in the Riemann tensor is measured only by the Ricci scalar $\Phi$, being equal to the local energy-momentum flux $T_{\mu \nu} \, l^{\mu} l^{\nu}$ across the null geodesics under consideration, where the vector field $l^{\mu}$ is everywhere tangent to the geodesic null congruence. Focusing effects over larger distances, however, which include both the Ricci scalar $\Phi$ and the Weyl scalar $\Psi$, may be captured by the inclusion of a mixed term similar to the mixed term in the total focusing power (\ref{focusingpower}) (see Condition ($\mathcal{O}$1)).

\subsection{Range of Validity} \label{SubsectionVB}

\noindent We are now in a position to discuss the limitations of Penrose's gravito-optical analogy, and thus to comment on its range of validity. For the sake of clarity, we here specify and discuss the limitations in form of a list:
\begin{itemize}
\item[($\mathcal{L}$1)] Considering spacetimes allowing for geodesic null congruences that exhibit nonvanishing rotations and/or dispersions, bundles of light may not behave like simple beams of rays, which is, however, required for the applicability of geometrical optics methods [($\mathcal{G}$1), ($\mathcal{G}$2), and ($\mathcal{G}$3)].
\item[($\mathcal{L}$2)] The interpretation of the (trace-free part of the) Ricci tensor as an anastigmatic lens and that of the Weyl tensor as an astigmatic lens are direct consequences of the particular modeling of both the Ricci and Weyl scalars $\Phi$ and $\Psi$ by means of Dirac delta distributions (which can be viewed as local gravitational lenses embedded in an otherwise flat spacetime) on the one hand, and of the assumption that these scalars are the dominant contributions in the Sachs equations on the other. However, since for arbitrary spacetimes the functional shapes of $\Phi$ and $\Psi$ in general differ from simple Dirac delta distributions, and since they do not constitute the dominant parts of the Sachs equations, this specific modeling (and direct extensions thereof) precludes the analysis of the effects of entirely different forms of spacetime curvature with deviating focusing properties. Accordingly, the above interpretations are not sustainable in general [($\mathcal{G}$4), ($\mathcal{G}$5), and ($\mathcal{G}$6)].
\item[($\mathcal{L}$3)] Only the Ricci and Weyl scalars $\Phi$ and $\Psi$, which do not account for the entire Ricci and Weyl tensors but merely certain components thereof (see Appendix A), are shown to exhibit, in the above limiting sense, the behaviors of anastigmatic and astigmatic lenses, respectively. Since the functional shapes of the remaining Ricci and Weyl scalars are left unspecified, the gravito-optical analogy cannot be employed for the full Ricci and Weyl tensors without further analysis [($\mathcal{G}$4)].
\item[($\mathcal{L}$4)] For spacetimes violating the null energy condition, which is, e.g., the case for Kerr--Vaidya black hole spacetimes \citep{DahalTerno}, a positive overall focusing via the Ricci scalar $\Phi$ is not possible [($\mathcal{G}$5)].
\item[($\mathcal{L}$5)] By only taking into account systems consisting of two thin, convex, astigmatic lenses that are parallel and of equal strengths, orthogonally-oriented, and described effectively on suitable scales, the analysis is restricted to those spacetimes where the focusing effect of the Weyl scalar $\Psi$ is essentially like that of a single thick, convex, anastigmatic lens, which corresponds to a Ricci-curved spacetime region [($\mathcal{O}$2)--($\mathcal{O}$4)].
\end{itemize}
Accordingly, as Penrose himself acknowledges [page 259], the range of validity of the gravito-optical analogy as it stands is, strictly speaking, limited to the particular class of spacetimes that behave like Dirac delta distributional gravitational lenses, comprise optical scalars satisfying the Sachs equations with the Ricci and Weyl scalars $\Phi$ and $\Psi$ being the leading contributions (where the former dominates the latter on scales much smaller than the radii of curvature involved in the Riemann tensor), obey the null energy condition, and allow for nondispersive, rotation-free geodesic null congruences. Hence, in its current form, the analogy is not general, and therefore does not directly apply to arbitrary spacetimes. We point out, however, that this does not imply that Penrose's original notion of energy cannot be suitably extended to hold for more general spacetimes. It simply means that the analogy between Ricci-curved and Weyl-curved spacetime regions on the one hand, and anastigmatic and astigmatic lenses on the other, has thus far been established to hold only under very specific circumstances. Therefore, the extent to which the above limitations can or cannot be relaxed determines whether the analogy can be considered beyond its given scope, as Penrose said he hoped it would be.

\subsection{Proof of Concept} \label{SubsectionVC}

\noindent In order to specify in what sense the focusing of curved spacetime regions and of optical lenses are analogous, we discuss Penrose's proof of concept of his observation that purely Ricci-curved Dirac delta regions of spacetime have an anastigmatic focusing effect on beams of light rays and purely Weyl-curved Dirac delta regions an astigmatic focusing effect [page 270]. For clarity, however, we here give a more detailed account of the proof. In the process, we also show that Penrose’s gravito-optical analogy has to be revised and weakened to the statement that purely Ricci-curved Dirac delta regions of spacetime influence light exactly like anastigmatic lenses, as he had originally argued, but that purely Weyl-curved Dirac delta regions act on light only similarly to---and not exactly like---astigmatic lenses. 

We begin by rewriting the two nontrivial rotation-free Sachs equations (\ref{ChangeOS1}) and (\ref{ChangeOS3}) (see Appendix B) in the form of a system
\begin{equation*}
D \boldsymbol{P} = \boldsymbol{P}^2 + \boldsymbol{Q} \, ,
\end{equation*}
with
\begin{equation*}
\boldsymbol{P} = \left(\begin{array}{cc}
\rho & \sigma \\
\overline{\sigma} & \overline{\rho} \\
\end{array}\right) 
\quad \textnormal{and} \quad 
\boldsymbol{Q} = \left(\begin{array}{cc}
\Phi & \Psi \\
\overline{\Psi} & \Phi \\
\end{array}\right) .
\end{equation*}
Employing the ansatz 
\begin{equation} \label{ansatz}
\boldsymbol{P} = - (D \boldsymbol{X}) \boldsymbol{X}^{- 1} \, ,
\end{equation}
this first-order, nonlinear, inhomogeneous system of ordinary differential equations in $\boldsymbol{P}$ transforms into the second-order, linear, homogeneous system in $\boldsymbol{X}$ given by
\begin{equation*}
D^2 \boldsymbol{X} = - \boldsymbol{Q} \boldsymbol{X} \, .
\end{equation*}
Now, considering the case introduced in Condition ($\mathcal{G}$6), namely that of an idealized Dirac delta distributional gravitational lens, i.e., 
\begin{equation} \label{DirLens}
\boldsymbol{Q} = \boldsymbol{Q}_{\delta} \, \delta(r) \quad \textnormal{with} \quad \boldsymbol{Q}_{\delta} := \left(\begin{array}{cc}
\Phi_{\delta} & \Psi_{\delta} \\
\overline{\Psi}_{\delta} & \Phi_{\delta} \\
\end{array}\right) ,
\end{equation}
and using $D = \partial_r$, the general solution of this system reads
\begin{equation*}
\boldsymbol{X}(r) + \boldsymbol{X}(0) \, r \, \boldsymbol{Q}_{\delta} H(r) = \boldsymbol{C}_0 + \boldsymbol{C}_1 r \, ,
\end{equation*}
where
\begin{equation*}
H(r) := \begin{cases}
\, 0 \quad \textnormal{for} \,\, r < 0 \\[0.2cm]
\, 1 \quad \textnormal{for} \,\, r \geq 0 
\end{cases}
\end{equation*}
is the usual Heaviside step function. The initial conditions $\boldsymbol{X}(0) = \1_{2 \times 2}$ and $D\boldsymbol{X}(0) = - \boldsymbol{P}(0) = - \boldsymbol{Q}_{\delta}$ give rise to the constants $\boldsymbol{C}_0 = \1_{2 \times 2}$ and $\boldsymbol{C}_1 = \boldsymbol{0}$, and thus to the particular solution
\begin{equation} \label{PartSol}
\boldsymbol{X} = \1_{2 \times 2} - r \, \boldsymbol{Q}_{\delta} H(r) \, .
\end{equation}
Substituting this particular solution into the ansatz (\ref{ansatz}), we find that it corresponds to the optical scalars
\begin{equation*}
\boldsymbol{P} = \boldsymbol{Q}_{\delta} H(r) \bigl(\1_{2 \times 2} - r \, \boldsymbol{Q}_{\delta} H(r)\bigr)^{- 1} \, .
\end{equation*}

In order to determine the focal points of a congruence of null geodesics described by such a solution of the Sachs equations, that is, the points where neighboring null geodesics cross and where the convergence or the shear tends to infinity, we may compute the zeros of the area of the cross section of the null congruence
\begin{equation} \label{area}
a = \pi \, \textnormal{det}(\boldsymbol{X}) \, .
\end{equation}
Since these focal points can be interpreted as focal points of a (thin) gravitational lens of the form (\ref{DirLens}), they provide information on the particular nature of focusing induced by the Ricci and Weyl tensors, respectively. Hence, substituting (\ref{PartSol}) into (\ref{area}), we obtain the area
\begin{equation} \label{area2}
a = \pi \bigl[1 - r \bigl(\Phi_{\delta} - |\Psi_{\delta}|\bigr) H(r)\bigr] \bigl[1 - r \bigl(\Phi_{\delta} + |\Psi_{\delta}|\bigr) H(r)\bigr]
\end{equation}
with the associated zero set 
\begin{equation*}
a^{- 1}(0) = \bigl\{\Phi_{\delta} \pm |\Psi_{\delta}| = r^{- 1} \, \big| \, r > 0\bigr\} \, .
\end{equation*}
The form of this zero set shows that in case of both nonvanishing pure Ricci curvature with $\Phi_{\delta} > 0$ and $|\Psi_{\delta}| = 0$ as well as nonvanishing pure Weyl curvature with $\Phi_{\delta} = 0$ and $|\Psi_{\delta}| > 0$, there is exactly one focal point behind the lens. The geometric nature of the respective focal points is, however, different, which can be seen directly from Equation (\ref{area2}). To be more precise, since for the case of pure Ricci curvature both bracketed factors vanish at the focal point, it represents an actual point. In the case of pure Weyl curvature, only the second factor vanishes, and as a consequence the corresponding focal point is a line segment. Accordingly, Penrose's gravito-optical analogy is exact for purely Ricci-curved spacetime regions, viz., they focus \textit{exactly like} anastigmatic lenses with one focal point. In contrast, for purely Weyl-curved spacetime regions the analogy has to be weakened to being only \textit{similar to}---and not exactly like---astigmatic lenses in the sense that instead of having \textit{two} focal lines they have only \textit{one} (cf.\ Section \ref{SubsectionIVB}). This aspect is, however, not of paramount importance because the effect of astigmatism is nonetheless present. We point out, though, that for the mixed case where both the Ricci curvature and the Weyl curvature do not vanish and $\Phi_{\delta} > |\Psi_{\delta}|$, there exist two line-spread focal points behind the lens. All this indicates that Penrose's motivation for the gravito-optical analogy may have originated in an analysis of the optical scalars and the Sachs equations relating the changes of the optical scalars to the Ricci and Weyl curvatures, where the convergence tending to infinity focuses the congruence of null geodesics into a point and the shear tending to infinity focuses it into a line (see FIG.\ 2 in Appendix B), yet making no connection to the actual number of focal points. This will be further elaborated in Section \ref{SectionVI}.

\subsection{Application to the Notion of Energy Flux in General Relativity} \label{SubsectionVD}

\noindent After imposing Conditions ($\mathcal{G}$1)--($\mathcal{G}$6) on gravitational systems, Conditions ($\mathcal{O}$1)--($\mathcal{O}$4) on optical lens systems, and applying the gravito-optical analogy in the form of Condition ($\mathcal{G}$7), Penrose studies his new concept of energy-momentum flux, which relates to the focusing power of spacetime curvature, in detail for three different physically relevant examples that he calls ``(approximately) plane gravitational waves,'' ``plane-polarized gravitational wave packets,'' and ``arbitrary systems.'' 

In the first example of ``(approximately) plane gravitational waves,'' for which only the Weyl curvature is nonvanishing, he assumes two gravitational waves of short duration, both consisting of a nonoscillatory pulse of equal strength but opposite orientation (in the sense of Condition ($\mathcal{O}$2)), which are located at fixed positions and separated by an affine distance $w$ on the null geodesic under consideration [page 266, third paragraph]. For $w \rightarrow 0$, the effects of the pulses cancel out and no energy is intercepted by the geodesic null congruence. However, for $0 < w p \ll 1$, where $p$ is the total focusing power of the two gravitational pulses, their effects---and hence their energy---appear to be essentially the same as that of a single ``matter pulse'' with total focusing strength $p = p_0^2 w$, $p_0$ denoting the focusing strength of each of the gravitational pulses, for which only the Ricci curvature is nonvanishing. This result is a direct consequence of the link between the particular optical system consisting of two thin, convex, astigmatic lenses and a single thick, convex, anastigmatic lens [page 264, last paragraph; page 266, up to Section 4] in combination with the gravito-optical analogy where the Weyl curvature is related to astigmatic lenses and the Ricci curvature to anastigmatic lenses [page 263, paragraph below Equation (19)]. 

In the second example, Penrose moves on to the more general case of a ``plane-polarized gravitational wave packet,''\footnote{He likewise conducts studies of electromagnetic as well as of zero-mass Dirac wave packets, which lead to similar results as in this case (except for a spin-related functional dependency of the affine distance in the total focusing power).} which he models as a series of $2 N$ nonoscillatory pulses, with $N \in \mathbb{N}$, intercepted by a geodesic null congruence [page 267, second paragraph]. Each of these pulses is again of equal strength, opposite to the preceding one, and separated by the same affine distance $w$ along the geodesic null congruence. Combining these particular pulses into pairs allows him to regard each pair as a single thick, convex, anastigmatic lens instead of two separate thin, convex, astigmatic lenses. Then, considering scales for which $w \sqrt{N} \, |\Psi_{\delta}| \ll 1$, where $\Psi_{\delta}$ is again the focusing strength of each pulse, so that the gravitational wave packet is sufficiently weak as well as locally confined, the total focusing power of the gravitational wave packet---and hence its total energy-momentum flux across the geodesic null congruence---can be obtained by simply summing up the focusing powers of all pairs resulting in $p \sim N w \, |\Psi_{\delta}|^2$.

Finally, to define the focusing power of a ``general system,'' which he regards as an arbitrary system of finite extent along the geodesic null congruence, i.e., the Ricci and Weyl scalars $\Phi$ and $\Psi$ vanish for values of the affine parameter outside of a certain finite range, he simply uses the same quantities as in the analysis of the focusing power of a system that can effectively be described by a single thick, convex, anastigmatic lens [page 269, second paragraph]. More precisely, by assuming that the Ricci scalar $\Phi$ is the dominant quantity in gravitational focusing, he defines the total energy-momentum flux across a geodesic null congruence up to some point $P$ by the focusing strength $\Phi_{\delta}$.\footnote{This quantity corresponds to the expression $\rho \xi + \sigma \overline{\eta}$ in Penrose's Equation (48) on page 271, which can be easily seen from his Equations (37) and (43) on pages 269 and 270, respectively.}

\section{Penrose's Gravito-optical Analogy in Light of Norton's Material Theory of Induction} \label{SectionVI}

\noindent We have done some quite heavy lifting by now. After Sections \ref{SectionII} and \ref{SectionIII} gave a historical account of the context and argumentative structure of Penrose's gravito-optical paper, Sections \ref{SectionIV} and \ref{SectionV} gave a fine-grained physical/mathematical analysis of the gravitational and optical systems Penrose analyzed, linked to each other, and used to provide a new way of defining energy-momentum flow in general relativity. Among other things, we amended Penrose's analogy to the result that Ricci-curved spacetime regions influence light passing through them \textit{exactly} like anastigmatic lenses, whereas Weyl-curved spacetime regions influence light only \textit{similarly} to astigmatic lenses.\footnote{See the end of Section \ref{SectionIII} and all of Section \ref{SubsectionVC}.}

This result already points to the possibility that the analogy is quite subtle: it is not enough to call it a perfect or an imperfect analogy, because it is perfect for one kind of spacetime curvature/lens and less perfect for the other. Furthermore, contrary to what many philosophical accounts of analogy would have you believe, in the details of how Penrose progresses there is no completely clear-cut distinction between a well-known source system on the one hand from which knowledge is transferred to a less well-known target system on the other. To be sure, at first sight Penrose clearly wants to transfer results from the domain of geometrical lens optics to the domain of gravitational systems as described in general relativity. But as we shall see in the following when analyzing \textit{the details} of how Penrose progresses, he goes back and forth between both sides of the analogy, taking inspiration from either side of the analogical divide to decide what he should do next on the other side. 

In reconstructing this back and forth, we shall use Norton's account of analogy as embedded in his material theory of induction. This will help us get closer to the bottom of Penrose's analogy, to better understand \textit{in what sense} some curved spacetime regions and some optical lenses are analogous, \textit{why} they are analogous, and \textit{how} the analogy can be used to derive new results. In order to do this properly, we first need to know a bit about Norton's material theory of induction.\footnote{It would also be interesting to attempt a reconstruction using the accounts of analogy of \citet{Hessebook} and \citet{bartha2010parallel}; however, we found Norton's account particularly pliable in analyzing Penrose's analogy.} 

In general, inductive inference may be construed in contrast to deductive inference. Whereas in the latter, true premises securely imply a true conclusion, this is not so for inductive inferences; here, the premises at best strongly \textit{support} the conclusion.\footnote{Eliminative induction can do better, but this special case of inductive reasoning does not seem to apply to the inferences discussed below.} The big question is how exactly this supporting works, and how one can distinguish between a successful/reliable inductive inference and an unsuccessful one. 

This is what Norton's account wants to answer in the analysis of particular historical case studies. In the preface of \citet[p.\ 7]{Norton2021materialinduction}, Norton points out that his account of induction rests on two core principles. Firstly, ``[a]ll induction is local,'' meaning that all inductive inferences are restricted to a certain domain, and that the reliability of inductive inferences in that domain depends on the \textit{facts} that have been established about this domain. Secondly, ``[t]here are no universal rules for inductive inference,'' i.e., there is no general inductive logic akin to deductive logic whose rules of inference apply to all domains equally. Instead, Norton argues, each domain of investigation will feature certain \textit{warranting facts}, and whether or not we tend to make reliable inductive inferences in said domains depends on whether we have a good handle on these warranting facts, which \textit{fuel} successful inductive inferences.

Two case studies that Norton discusses in detail are Marie Curie's inference towards the crystalline structure of radium, and Galileo's inference that the patterns he saw on the Moon when looking through a telescope were mountains throwing shadows. With regard to the Curie case, Norton notes that it would be all too easy to see what Curie did as a special case of the inference schema ``Some samples of radium have been found to have certain chemical properties. Thus, all samples of radium have these properties,'' and based on that ``Some samples of radium have been found to have chemical properties similar to those of barium. Barium has this particular crystalline structure. Thus, all samples of radium will have a very similar crystalline structure.'' Put like this, it looks like fallacious deductive reasoning or very naive inductive reasoning, prone to fail in a heartbeat. Norton's point is that analyzing Curie's reasoning in this way is just not doing her justice, and that her actual reasoning rests on long-established \textit{facts} that \textit{drive} her analysis and \textit{warrant} the inductive inference that eventually got her the Nobel Prize for Chemistry in 1911 for the discovery and analysis of radium and polonium.\footnote{Of course, Curie had already been awarded the Nobel Prize in Physics for the development of a theory of radioactivity in 1903, making her the only person who ever won the Nobel Prize in two distinct scientific fields. For an analysis of Curie's work on radium in the context of the material theory of induction see \citet[Section 1.9]{Norton2021materialinduction}.} In a nutshell, the fact that had been established in chemistry research of the late 18th and early 19th century was that ``Generally, salts that have similar chemical properties have similar crystalline structure.''\footnote{This is a slightly different way of putting it than Norton does on page 46 of \citet{Norton2021materialinduction}; he speaks of salts being ``chemically analogous'' instead of them having ``similar chemical properties.''} It is the strength of evidence that has been found for this general rule---which \textit{does} allow for exceptions---that fuels Curie's analysis and allows her to make a successful inductive inference. 

The Galileo case, however, is much closer to our own example. Norton argues that Galileo made an \textit{inference by analogy}, which he takes to be a special case of an inductive inference. Referring to \citet{Mill:1904} and \citet{Joyce:1936}, he notes that there has been a long tradition of formalizing analogical inferences as inferences of the form ``$A$ is $P$. $B$ resembles $A$ in being $M$. Thus, $B$ is $P$.'' Norton regards this one-fits-all way of formalizing analogical inferences as problematic; but he also points out that Joyce noted that the reliability of the inference depends on there being a ``causal connection'' between $P$ and $M$, which is much more in the direction in which he develops his own account. In the latter, Norton replaces Joyce's demand by the requirement for there to be a factual connection, a ``warranting fact to authorize'' the analogical inference \citep[][p.\ 120]{Norton2021materialinduction}. And if such a fact of analogy has been conjectured, or even explored or established empirically, then this ``warrants an analogical inference, the passing of particular properties from the source system to the target'' (p.\ 131). Norton emphasizes that it is crucial to distinguish the original ``fact of analogy'' on the one hand, and the ``analogical inference warranted by a fact of analogy on the other.'' The former is a crucial \textit{premise} that needs to be established before any analogical inference can be made, for it to then serve as an (often implicit) premise in the inductive argument. In contrast, the latter is the \textit{inference itself} that leads to a particular conclusion about the target system.

All this plays out well in Norton's analysis of Galileo's inference to there being mountains on the Moon, reported in Galileo's 1610 ``Siderius Nuncius.'' Norton argues that when observing the Moon with the then novel telescope, Galileo did not see mountains on the Moon directly, but instead inferred their existence from what he did see:\footnote{Cf.\ \citet[p.\ 134]{Norton2021materialinduction}.} 

\begin{quote}
He tracked the advancing division between light and dark on the waxing Moon. His telescope showed that its edge was not a smooth curve but an ``uneven, rough and very wavy line.'' More important was the way it changed over time. As it slowly advanced, bright points of light would appear ahead of it. They would grow and soon join up with the advancing edge. Galileo found the analogy to the illumination of mountains on Earth irresistible.
\end{quote}

But what was the ``fact of analogy'' that fueled Galileo's inference? Norton argues that it is the previously established fact that the Moon is the same kind of object as the Earth, and that light hitting elevations on the Moon would react in just the same way (analogously) as it reacts when it hits an elevation on Earth: it is reflected when the elevation is solid, and bent and refracted on the edges of the object. In a word: light will make the elevation on the Moon throw a shadow, just as it does on Earth.\footnote{We said above that the fact of analogy could have been established or merely conjectured. It is fair to say that in Galileo's case it was merely conjectured, and that indeed such a conjecture was very bold and unusual at the time. For since Plato and Aristotle at least it had \textit{instead} been conjectured that the heavenly bodies, including the Moon, were made of an entirely different element than the Earth and the objects on Earth, and that they behaved according to a separate set of laws of nature.} On the basis of this fact and using the laws of optics, Galileo could then \textit{deductively} show how the advancing division between light and dark would move if they were actually shadows thrown by mountains, and he could even calculate the height of these mountains. Crucially, the fact of analogy that fuels the inference is not some brute similarity between the Earth and the Moon as such, but that they are both specific examples of objects with uneven surfaces that turn while being under unidirectional light and thus throw shadows, and that the laws of optics (or simply shadow casting) apply to both Earth and Moon in the same way. 

We believe that Penrose's analogy between curved spacetime regions and optical lenses functions in a very similar way. The similarity is less obvious than ``mountains here interact with light the same way as mountains there.'' But the comparison between Galileo's analogy and Penrose's analogy allows us to see a (meta-)similarity: When it comes down to it, both curved spacetime regions and optical lenses are examples of transparent media through which light moves, and in which its path is influenced by the curvature of the medium.\footnote{Of course, this does not mean that there are no important differences between curved spacetime regions on the one hand, and optical lenses on the other, even when it comes to them acting as media for light. For example, optical lenses are in general homogeneous media that do not vary in their density, whereas curved spacetime regions do not have a density in the same sense; and yet, the curvature of a spacetime region can influence light in ways akin to the curvature of an optical lens.} Of course, it could have been that a curved spacetime region is curved in a very different way from a curved optical lens. In fact, Penrose has \textit{not} established that for every way that spacetime can be curved there is a corresponding optical lens that is curved just so it would influence a light beam passing through in the same way as the curved spacetime region. Instead, he did something much more specific. Most likely being inspired by Sachs' invention of the optical scalars in general relativity,\footnote{See Section \ref{SectionII} for the historical reconstruction and context, and Appendix B for the mathematically precise account.} Penrose likely saw that nonvanishing convergence $\theta$ influences geodesic null congruences quite similarly to how anastigmatic lenses focus light rays, and that nonvanishing shear $\sigma$ influences these congruences very much like astigmatic lenses affect light. 

But in order to see how far this analogy could be pushed, Penrose needed to look for gravitational systems that behave like anastigmatic and astigmatic lenses as much as possible: in Norton's words, he needed to find or construct systems that would allow nailing down the analogical fact that would authorize analogical inferences about gravitational systems. For as he himself stated in his introduction, that was his aim: infer something new about gravitational systems, and especially about gravitational energy, by comparing gravitational systems to optical lenses. 

Compared to Galileo, this was a much harder task. Galileo inferred that mountains on the Moon exist, just like mountains on Earth exist---but almost everything else followed after that: Moon mountains interact with light the same way as Earth mountains; they are made from essentially the same material, or so it was plausible to assume. In Penrose's case, the material of the medium through which light moves is not the same: in one case the material is glass (or plastic), in the other curved spacetime. However, Penrose clearly thought that it was worthwhile to look for spacetimes that treat light just like optical lenses do; Penrose was searching for the analogical fact that would warrant the analogical inferences he hoped to make. 

As we saw in Section \ref{SectionIII}, Penrose starts out by introducing Sachs' optical scalars, convergence $\theta$, rotation $\omega$, and shear $\sigma$ in his Section 2, ``Optical Scalars.'' Having gone through the whole paper itself, and the parts of general relativity and lens optics it implicitly refers to, it now seems plausible that this is also the origin of what was likely initially only a hunch that there might be an analogy between certain gravitational systems and certain optical systems. Indeed, from Figure 1 in Penrose's paper (reproduced as FIG.\ 2 in our Appendix B), which visualizes his interpretation of the optical scalars, one might get the idea that the convergence $\theta$ focuses a null congruence perfectly like an anastigmatic lens focuses a beam of light rays, and that the shear $\sigma$ blurs a null congruence in a way quite similar to what an astigmatic lens does to a beam of light rays.\footnote{Our conjecture of the temporal order of hunches, ideas, and developed reasoning is further strengthened by Penrose's 1967 Battelle Lectures. In \citet[pp.\ 166--168]{penrose-struc-st2}, Penrose first reproduces the same figure of the optical scalars, and then argues that the part of the Sachs equations governing convergence $\theta$ and shear $\sigma$ can be interpreted as showing the analogy between Ricci-curved regions of spacetime to anastigmatic lenses and of Weyl-curved regions to astigmatic lenses, respectively. Thus, his order of presentation corresponds to our reconstructed order of discovery/reasoning.} 

But as so often, a hunch is not enough, especially if you want to use an analogy as a warranting fact on the basis of which to make inferences. In the following, we shall use Norton's terminology to show that Penrose's paper can be understood as his first conjecturing, then establishing, and finally rendering more precisely a fact of analogy between gravitational and optical systems. We shall also see that Penrose is using an iterative process to do so, going back and forth between gravitational and optical systems in his construction, in a way buttressing the analogy on both sides of the fence. 

The endeavor starts shortly before his Section 3, ``Lenses,''\footnote{Recall our short summary of this section in our Section \ref{SectionIII}.} with his establishing what we called Conditions ($\mathcal{G}$1)--($\mathcal{G}$3) in our Section \ref{SubsectionVA}. These three conditions that Penrose imposes on the gravitational systems have the aim of making the null congruences to be investigated very much like beams of light rays in geometrical optics. Condition ($\mathcal{G}$1) restricts the measurement of focusing power of curved spacetime regions to along geodesic null congruences, just like the focusing power of optical lenses is measured along light rays. Condition ($\mathcal{G}$2) restricts the null congruences to be investigated to those that have vanishing rotation $\omega$, for the light rays of geometrical optics cannot twist. Condition ($\mathcal{G}$3), finally, demands that the null congruences are nondispersive, for the optical phenomenon of dispersion cannot be handled in geometrical optics, ray optics, but only in wave optics. 

So far so good; but in order to connect the convergence and shear of a geodesic, rotation-free, nondispersive null congruence to the Ricci and Weyl curvatures of the spacetime the congruence moves through, Penrose needed to take a closer look at the Sachs equations (\ref{ChangeOS1}) and (\ref{ChangeOS3}) that describe their development through spacetime. Having previously derived the Ricci and Weyl scalars in his work with Newman \citep{NewmanPenrose1962}, Penrose could now at the beginning of his Section 3 argue that the convergence $\theta$ is dominated by the Ricci scalar $\Phi$, whereas the shear $\sigma$ is dominated by the Weyl scalar $\Psi$. Further restricting the set of null congruences to those for which this is the case is done by our Condition ($\mathcal{G}$4). Thus, a spacetime region which makes a null congruence primarily converge must be mostly Ricci-curved, whereas a spacetime region that primarily produces shear in a null congruence must be mostly Weyl-curved. In the case of a Ricci-curved region, demanding that it makes the null geodesics converge to a single point like an ideal magnifying glass requires also imposing the null energy condition, Condition ($\mathcal{G}$5), on the Ricci tensor.

Thus far, the conditions imposed on gravitational systems have not yet brought about a counterpart to optical lenses. Penrose models the corresponding spacetime regions by imposing Condition ($\mathcal{G}$6), the condition that specifies the Ricci and Weyl scalars in the spacetime region as Dirac delta distributions; so all the curvature that is felt by the null congruence is concentrated at a $3$-dimensional hypersurface that intersects the path of the null congruence through spacetime, just like the optical lens intersects the path of the light ray through space.

Imposing Conditions ($\mathcal{G}$1)--($\mathcal{G}$6) has then specified two types of spacetime regions (purely Ricci-curved Dirac delta regions and purely Weyl-curved Dirac delta regions) that act on geodesic, rotation-free, nondispersive null congruences very much like anastigmatic and astigmatic lenses, respectively, act on beams of light rays. While Galileo could assume that mountains on the Moon react to light in the same way as mountains on Earth do because they are made of the same material, Penrose had to find, or even build, those gravitational systems in the solution space of the Einstein field equations that would react to light akin to these two types of optical lenses. Having found these systems, he thus \textit{established} a fact of analogy between particular gravitational systems and particular optical lenses. At this point in the construction, it is an established fact, as Penrose himself also states at the beginning of his Section 3, that beams of light rays (from now on identified with geodesic null congruences) passing through a spacetime region governed by Conditions ($\mathcal{G}$1)--($\mathcal{G}$6) are influenced in a way analogous to light passing through anastigmatic and astigmatic lenses. 

Now that Penrose has well-defined gravitational systems, he can explore their properties independently of the analogy that motivated the conditions that gave rise to them. It is as if Galileo had teamed up with the crew of Apollo 11 and asked them to check whether the mountains on the Moon are \textit{really} made from the same stuff as the mountains on Earth. But to really get to the bottom of this question, Armstrong, Aldrin, and Collins would have had to bring back rock samples from the Moon and compare them to rock samples from Earth.

Likewise, the sought-after gravitational systems in hand, and having investigated their properties further, Penrose now turns around and checks whether certain properties he found in the gravitational domain could be associated with corresponding properties in the optical domain. Thus, Penrose not only imposes conditions on the spacetime regions in question but also does the same to the optical systems under consideration, supposed to be analogous. Indeed, Condition ($\mathcal{O}$1) from our Section \ref{SubsectionVA} makes sure that the analogy is restricted just as much in the optical domain, and not only to anastigmatic and astigmatic lenses as such, but to those subsets of these two classes of lenses that feature properties corresponding to the gravitational systems previously identified.\footnote{At this point, we are still within our initial summary of Penrose's Section 3, given in our Section \ref{SectionIII} on page \pageref{Psec3Intro}.} In particular, Penrose looks for an optical counterpart to the nonlinear features of his gravitational systems exhibited by the nonlinearity of the Sachs equations, finds it in the nonlinear contribution to the total focusing power of optical systems consisting of two thin, convex, anastigmatic lenses, and then turns around yet again to transfer this to the associated gravitational systems (see again Condition ($\mathcal{O}$1) and Condition ($\mathcal{G}$7)). 

We will come back to this in a moment; but let us first note that this type of iterative process, this going back and forth and transferring properties \textit{both} from the source system to the target system and vice versa in an iterative fashion, is not present in Norton's analysis of Galileo (or Curie). But we believe it is not atypical for the use of analogies in physics. The iterative process allows one to make the analogy ever more precise. It is just like an artist painting a portrait: the artist is primarily in the business of creating a picture, her target, that corresponds to a source, the model. But in between painting different parts of the model, she might focus on particular aspects of the model, inspired by her own act of painting, aspects that she would not have focused on otherwise.

Similarly, Penrose clearly wants to find a certain set of spacetime regions/gravitational systems---that is his target. But while searching and investigating and then polishing the target of his labors, he occasionally looks back at the source system, the optical lenses, and zooms in on particular aspects, inspired by what he might next need for the gravitational domain. 

Having imposed Condition ($\mathcal{O}$1) on the optical systems (and for reasons of mathematical simplicity Conditions ($\mathcal{O}$2)--($\mathcal{O}$4)), all in light of Conditions ($\mathcal{G}$1)--($\mathcal{G}$6), Penrose then turns around yet again and returns to the gravitational systems. He finds that the newly imposed conditions ($\mathcal{O}$1)--($\mathcal{O}$4) force him to set one more condition on the gravitational systems: Condition ($\mathcal{G}$7). This is the strongest transfer yet, as it allows him to carry the concept ``focusing power'' over from the domain of optical lenses (subject to Conditions ($\mathcal{O}$1)--($\mathcal{O}$4)) to the domain of gravitational systems (subject to Conditions ($\mathcal{G}$1)--($\mathcal{G}$6)). And not only that, learning from lens optics he is able to delineate \textit{precisely} under which conditions the focusing power of gravitational systems is additive and when it is not. This then brings us, at the beginning of Penrose's Section 4, to the controlling fact of the analogy, the core or the correspondence relationship that Penrose was seeking. Having found that spacetime regions that are curved in a particular way influence light rays analogously to two particular types of optical lenses, Penrose finally turns to a fact that had been established way before and that would enable him to close the bridge between gravitational and optical systems: the Einstein field equations. As he had identified total focusing power locally with Ricci curvature, and given that the Einstein field equations in their trace-reversed form (\ref{Einstein}) identify Ricci curvature with energy-momentum, he could finally identify the total focusing power, due to Ricci and/or Weyl curvature, of a spacetime region with the energy-momentum flux through that region. This, then, is the controlling fact for all analogical inferences that follow: that under certain conditions the energy-momentum flux through a spacetime region can be obtained from its focusing power, i.e., its influence on null congruences. Based on this, Penrose can now use the fact of analogy to authorize analogical inferences, and thus make predictions about gravitational waves and energy-momentum flux in his Section 4. Just like in Galileo's case and unlike in Curie's case,\footnote{See \citet{Norton2021materialinduction}, Chapter 1, p.\ 40, for Curie and Chapter 4, p.\ 134, for Galileo in this regard. Note that in the case of Curie, her predecessor Huay had developed the theory of crystals in such a way that after the warranting fact had been established, things proceeded deductively from there. However, in this case this was too restrictive and ultimately unsuccessful \citep[cf.][p.\ 42]{Norton2021materialinduction}, so that Curie's fully inductive account superseded Huay's.} Penrose's analogy is what Norton called one of these rare cases in which ``inductive inferences may turn out to have been deductive inferences all along, once we make the background facts explicit [...] an extreme and relatively rare case;'' though the background facts will still have been established inductively \citep[][p.\ 51]{Norton2021materialinduction}.  Our expectation is that in the realm of physics, such cases may not be all that rare; and as Norton points out in a footnote to the above quote, there is still plenty of inductive risk in the establishment of the facts of analogy that serve as a premise in the subsequent deductive arguments.

Just like the iterative process described above, Penrose's Section 5 is something that does not appear in Norton's discussion of Galileo and Curie, but which is nonetheless compatible with Norton's account. Having first conjectured and then established a (series of) fact(s) of analogy between specific curved spacetime regions and specific optical lenses, Penrose now wants to see how far the analogy can be pushed, whether it can be extended to more general spacetime regions. In order to do this, he has to generalize and also to make more precise his original argument for the correspondence, and in our reconstruction of the generalized argument (see Section \ref{SubsectionVC} for the proof of the original argument and the end of Section \ref{SubsectionVD} for the generalized argument), one finds not only the precise range of validity for both of Penrose's arguments (cf.\ Section \ref{SubsectionVB}), but that already within the range of the original argument, there is an interesting wrinkle in the analogy. It is not simply that the analogy is not perfect---it is perfect for Ricci-curved Dirac delta regions and anastigmatic lenses but imperfect for Weyl-curved Dirac delta regions and astigmatic lenses. To say it one more time: Ricci-curved Dirac delta regions behave \textit{exactly} like anastigmatic lenses, whereas Weyl-curved Dirac delta regions only behave \textit{similarly} to astigmatic lenses. Furthermore, the link between focusing power and energy-momentum flux \textit{really} works only for how Ricci-curved spacetime regions influence null congruences; the link works only for those Weyl-curved regions that can be modeled by Ricci-curved regions.\footnote{See Section \ref{SubsectionVB}.}

In our opinion, though, this only makes the analogy conceptually more intriguing. For the question Penrose asked at the beginning of his paper still stands to this day: how far can the analogy be extended, to an analogy between \textit{other} optical lenses and \textit{otherwise} curved spacetime regions? Even as it stands, it seems clear to us that the predictive potential of the analogy has not yet been exhausted. The fact of analogy that Penrose conjectured, established, generalized, and made more precise than it originally was may well authorize a whole plethora of analogical inferences that are yet to be made. And what type of analogical inferences would they be? 

Let us take stock. At the beginning of our paper, we asked whether Penrose's analogy was a metaphor of only heuristic value, maybe even just a visual and quite possibly misleading aid, or whether it encoded something deeper about general relativity as such and the nature of spacetime curvature in particular. We can now say with conviction that it is indeed more than the metaphor of a curved sheet of cloth---much more. It does not \textit{only} allow one to visualize particularly curved spacetime regions as optical lenses of a certain kind; it allows one to derive new and hitherto unfathomed results about such spacetime regions, and even to bring about a new---and well-defined---notion of energy in general relativity.

\vspace{0.2cm}

\section*{Acknowledgments}

\noindent The authors gratefully acknowledge funding from the European Research Council, Grant 101088528 COGY, from the Lichtenberg Grant for Philosophy and History of Physics of the Volkswagen Foundation, and from the German Research Foundation, DFG Research Group FOR 2495. In addition, JD acknowledges the support of the Black Hole Initiative, Harvard University. We are also grateful to the Lichtenberg Group for History and Philosophy of Physics of the University of Bonn for useful discussions and comments on the first draft of this paper. We are particularly grateful to Erik Curiel and Samuel Fletcher for detailed and careful comments on a later draft, and to the two anonymous referees for their constructive recommendations and suggestions.

\vspace{0.2cm}

\begin{appendix}

\section*{Appendix A: Newman--Penrose Formalism and Spinors} 

\noindent We give a short account of all the aspects of the Newman--Penrose formalism \citep{NewmanPenrose1962} required for Penrose's gravito-optical analogy, which includes its $2$-spinor representation. To this end, we first recall the more general tetrad formalism. We again let $(\mathfrak{M}, \boldsymbol{g})$ be a Lorentzian $4$-manifold endowed with the Levi-Civita connection $\boldsymbol{\nabla}$. Furthermore, we introduce two types of dual pairs of bases for the tangent and cotangent spaces $T_p\mathfrak{M}$ and $T^{\star}_p\mathfrak{M}$ at each point $p \in \mathfrak{M}$, namely dual pairs of coordinate bases $(\boldsymbol{e}_{\mu})$ and $(\boldsymbol{e}^{\mu})$ as well as dual pairs of tetrad bases $(\boldsymbol{e}_{(a)})$ and $(\boldsymbol{e}^{(a)})$ consisting of frame fields, i.e., four locally defined, linearly independent, orthonormal or null vector fields, and their unique metric duals. Here, Greek letters label the tensor indices and Latin letters enclosed in parentheses label tetrad indices. Fixing a dual pair of coordinate bases, a dual pair of tetrad bases can be expressed as
\begin{equation*}
\boldsymbol{e}\indices{_{(a)}} = e\indices{_{(a)}^{\mu}} \, \boldsymbol{e}\indices{_{\mu}} \quad \textnormal{and} \quad \boldsymbol{e}\indices{^{(a)}} = e\indices{^{(a)}_{\mu}} \, \boldsymbol{e}\indices{^{\mu}} \, , 
\end{equation*}
where $e\indices{_{(a)}^{\mu}} \hspace{-0.07cm} : T\mathfrak{M} \rightarrow T\mathfrak{M}$ are linear mappings with $e\indices{_{(a)}^{\mu}} \, e\indices{^{(b)}_{\mu}} = \delta^{(b)}_{(a)}$ and $e\indices{_{(a)}^{\mu}} \, e\indices{^{(a)}_{\nu}} = \delta^{\mu}_{\nu}$. These basis vectors are usually chosen in such a way that the inner product
\begin{equation*}
\boldsymbol{g}(\boldsymbol{e}_{(a)}, \boldsymbol{e}_{(b)}) = e\indices{_{(a)}^{\mu}} \, e\indices{_{(b) \, \mu}} = \eta_{(a) \, (b)} 
\end{equation*}
yields a specified nondegenerate, constant, symmetric matrix $\boldsymbol{\eta} \in \textnormal{Sym}_{4}(\mathfrak{M})$, which serves as the metric in the tetrad formalism. Moreover, one has the freedom to perform both local Lorentz transformations $\Lambda\indices{^{(a)}_{(a')}}(x^{\mu}) \in O(1, 3)$ as well as general coordinate transformations $\partial_{\mu} x^{\mu'} \in \textnormal{Cov}(\mathfrak{M})$, where $\textnormal{Cov}(\mathfrak{M})$ denotes the covariance group of $\mathfrak{M}$, that is, a (pseudo-)subgroup of $\textnormal{Diff}^{1}(\mathfrak{M})$, the group of all $C^1$-diffeomorphisms of $\mathfrak{M}$ to itself. The Riemann tensor in this framework is defined via the (torsion-free) Maurer--Cartan equations of structure 
\begin{equation*}
\begin{split}	 
& \textnormal{d}\boldsymbol{e}^{(a)} + \boldsymbol{\omega}\indices{^{(a)}_{(b)}} \wedge \boldsymbol{e}^{(b)} = 0 \\[0.2cm]
& \textnormal{d}\boldsymbol{\omega}\indices{^{(a)}_{(b)}} + \boldsymbol{\omega}\indices{^{(a)}_{(c)}} \wedge \boldsymbol{\omega}\indices{^{(c)}_{(b)}} = \tfrac{1}{2} \, R\indices{^{(a)}_{(b) \, (c) \, (d)}} \, \boldsymbol{e}^{(c)} \wedge \boldsymbol{e}^{(d)} \, ,
\end{split}
\end{equation*}
in which the spin connection $1$-forms 
\begin{equation*}
\boldsymbol{\omega}\indices{^{(a)}_{(b)}} = \bigl(e\indices{^{(a)}_{\nu}} \, e\indices{_{(b)}^{\lambda}} \, \Gamma^{\nu}_{\mu \lambda} -  e\indices{_{(b)}^{\lambda}} \, \partial_{\mu} e\indices{^{(a)}_{\lambda}}\bigr) \, \textnormal{d}x^{\mu} 
\end{equation*}
are the tetrad formalism representation of the Christoffel symbols $\Gamma^{\nu}_{\mu \lambda}$. 

Now, the Newman--Penrose formalism is a specific tetrad formalism, in which the local tetrad bases consist of two real-valued null vectors, $\boldsymbol{l} = \boldsymbol{e}\indices{_{(0)}}$ and $\boldsymbol{n} = \boldsymbol{e}\indices{_{(1)}}$, as well as a complex-conjugate pair of null vectors, $\boldsymbol{m} = \boldsymbol{e}\indices{_{(2)}}$ and $\boldsymbol{\overline{m}} = \boldsymbol{e}\indices{_{(3)}}$, satisfying the null, orthogonality, and cross-normalization conditions
\begin{equation*} 
\begin{split}
& \boldsymbol{l} \cdot \boldsymbol{l} = \boldsymbol{n} \cdot \boldsymbol{n} = \boldsymbol{m} \cdot \boldsymbol{m} = \boldsymbol{\overline{m}} \cdot \boldsymbol{\overline{m}} = 0 \\[0.2cm]
& \boldsymbol{l} \cdot \boldsymbol{m} = \boldsymbol{l} \cdot \boldsymbol{\overline{m}} = \boldsymbol{n} \cdot \boldsymbol{m} = \boldsymbol{n} \cdot \boldsymbol{\overline{m}} = 0 \\[0.2cm]
& \boldsymbol{l} \cdot \boldsymbol{n} = - \boldsymbol{m} \cdot \boldsymbol{\overline{m}} = 1 \, . 
\end{split}
\end{equation*}	
The metric in this formalism reads
\begin{equation*}
\boldsymbol{\eta} = \boldsymbol{l} \otimes \boldsymbol{n} + \boldsymbol{n} \otimes \boldsymbol{l} - \boldsymbol{m} \otimes \boldsymbol{\overline{m}} - \boldsymbol{\overline{m}} \otimes \boldsymbol{m} \, .  
\end{equation*}
Moreover, the spin connection is represented by the spin coefficients 
\begin{eqnarray*}  
\kappa = \gamma\indices{_{(2) \, (0) \, (0)}} & \quad \varrho = \gamma\indices{_{(2) \, (0) \, (3)}} & \quad \epsilon = \tfrac{1}{2} (\gamma\indices{_{(1) \, (0) \, (0)}} + \gamma\indices{_{(2) \, (3) \, (0)}}) \nonumber \\
\sigma = \gamma\indices{_{(2) \, (0) \, (2)}} & \quad \mu = \gamma\indices{_{(1) \, (3) \, (2)}} & \quad \gamma = \tfrac{1}{2} (\gamma\indices{_{(1) \, (0) \, (1)}} + \gamma\indices{_{(2) \, (3) \, (1)}}) \nonumber \\
\lambda = \gamma\indices{_{(1) \, (3) \, (3)}} & \quad \tau = \gamma\indices{_{(2) \, (0) \, (1)}} & \quad \alpha = \tfrac{1}{2} (\gamma\indices{_{(1) \, (0) \, (3)}} + \gamma\indices{_{(2) \, (3) \, (3)}}) \\
\nu = \gamma\indices{_{(1) \, (3) \, (1)}} & \quad \pi = \gamma\indices{_{(1) \, (3) \, (0)}} & \quad \beta = \tfrac{1}{2} (\gamma\indices{_{(1) \, (0) \, (2)}} + \gamma\indices{_{(2) \, (3) \, (2)}}) \nonumber \, ,
\end{eqnarray*}
where the symbols $\gamma_{(a) (b) (c)}$ are the so-called Ricci rotation coefficients defined by
\begin{equation*}
\gamma\indices{^{(a)}_{(b) (c)}} \boldsymbol{e}\indices{^{(c)}} = e\indices{_{\mu}^{(a)}} \bigl[\textnormal{d}e\indices{^{\mu}_{(b)}} + \omega\indices{^{\mu}_{(b) (c)}} \boldsymbol{e}^{(c)}\bigr] \, .
\end{equation*}
The two central quantities in Penrose's analogy, namely the Ricci tensor $R_{(a) (b)}$ and the Weyl tensor $C_{(a) (b) (c) (d)}$, are expressed through the four real-valued scalars
\begin{align} \label{NPRicci}
& \Phi_{0 0} = - \tfrac{1}{2} \, R_{(0) (0)} = - \tfrac{1}{2} \, R_{\mu \nu} \, l^{\mu} l^{\nu} \\[0.2cm]
& \Phi_{1 1} = - \tfrac{1}{4} \, (R_{(0) (1)} + R_{(2) (3)}) = - \tfrac{1}{4} \, R_{\mu \nu} \, (l^{\mu} n^{\nu} + m^{\mu} \overline{m}^{\nu}) \nonumber \\[0.2cm]
& \Phi_{2 2} = - \tfrac{1}{2} \, R_{(1) (1)} = - \tfrac{1}{2} \, R_{\mu \nu} \, n^{\mu} n^{\nu} \nonumber \\[0.2cm]
& \Lambda = \tfrac{1}{12} \, (R_{(0) (1)} - R_{(2) (3)}) = \tfrac{1}{12} \, R_{\mu \nu} \, (l^{\mu} n^{\nu} - m^{\mu} \overline{m}^{\nu}) \nonumber
\end{align}	
and the three complex scalars
\begin{equation*} 
\begin{split}
& \Phi_{0 1} = - \tfrac{1}{2} \, R_{(0) (2)} = - \tfrac{1}{2} \, R_{\mu \nu} \, l^{\mu} m^{\nu} \\[0.2cm]
& \Phi_{0 2} = - \tfrac{1}{2} \, R_{(2) (2)} = - \tfrac{1}{2} \, R_{\mu \nu} \, m^{\mu} m^{\nu} \\[0.2cm]
& \Phi_{1 2} = - \tfrac{1}{2} \, R_{(1) (2)} = - \tfrac{1}{2} \, R_{\mu \nu} \, n^{\mu} m^{\nu}
\end{split}
\end{equation*}	
on the one hand, and through the five complex scalars 
\begin{align} \label{NPWeyl}
& \Psi_0 = - C_{(0) (2) (0) (2)} = - C_{\mu \nu \alpha \beta} \, l^{\mu} \, m^{\nu} \, l^{\alpha} \, m^{\beta} \\[0.2cm]
& \Psi_1 = - C_{(0) (1) (0) (2)} = - C_{\mu \nu \alpha \beta} \, l^{\mu} \, n^{\nu} \, l^{\alpha} \, m^{\beta} \nonumber \\[0.2cm]
& \Psi_2 = - C_{(0) (2) (3) (1)} = - C_{\mu \nu \alpha \beta} \, l^{\mu} \, m^{\nu} \, \overline{m}^{\alpha} \, n^{\beta} \nonumber \\[0.2cm]
& \Psi_3 = - C_{(0) (1) (3) (1)} = - C_{\mu \nu \alpha \beta} \, l^{\mu} \, n^{\nu} \, \overline{m}^{\alpha} \, n^{\beta} \nonumber \\[0.2cm]
& \Psi_4 = - C_{(1) (3) (1) (3)} = - C_{\mu \nu \alpha \beta} \, n^{\mu} \, \overline{m}^{\nu} \, n^{\alpha} \, \overline{m}^{\beta} \nonumber
\end{align}	
on the other hand. Further details on the tetrad formalism and the Newman--Penrose formalism can be found in, for example, \citet[Chapters 1.7. and 1.8.]{ChandraBook}.

We next present a compact representation theoretic/geometric account of spinors and the $2$-spinor representation of the Newman--Penrose formalism \citep[see, e.g.,][Chapter 2]{geroch1968spinor, PenroseRindler,LawsonMichelsohn}. Working within the framework of general relativity, we may consider the group $\textnormal{SL}(2, \mathbb{C}) \cong \textnormal{Spin}(1, 3) \rightarrow \textnormal{SO}^+(1, 3, \mathbb{R})$ as the local spinor group acting on the $2$-dimensional complex vector space $\mathbb{C}^2$ and preserving its natural $2$-dimensional nondegenerate skew-symmetric $2$-form $\boldsymbol{\epsilon}$, which is defined by the $2$-dimensional Levi-Civita symbol. Hence, we regard $(\mathbb{C}^2, \boldsymbol{\epsilon})$ as our spinor space and elements $\boldsymbol{\xi} \in \mathbb{C}^2$ as spinors. For each spinor $\boldsymbol{\xi}$ there exists a dual with respect to $\boldsymbol{\epsilon}$, namely a mapping $\mathbb{C}^2 \rightarrow (\mathbb{C}^2)^{\star}$ with $\xi^A \mapsto \xi_A = \epsilon_{B A} \xi^B$, where $A, B \in \{0, 1\}$. Additionally, since $\mathbb{C}^2$ is naturally endowed with a conjugacy operation, we can define a complex conjugate and a complex conjugate dual (denoted by a bar over the symbol with simultaneous priming of sub- and superscript letters). Then, in order to set up the $2$-spinor representation of the Newman--Penrose formalism, we introduce a local dyad basis $(\zeta_{(k)})$, $k \in \{1, 2\}$, for the spinor space $(\mathbb{C}^2, \boldsymbol{\epsilon})$ and a local dyad co-basis $(\zeta^{(k)})$ with respect to the $2$-dimensional Levi-Civita symbol $\boldsymbol{\epsilon}$ for its dual. Projecting the spinor $\boldsymbol{\xi}$ onto this dyad basis and the dual spinor onto the dyad co-basis yields
\begin{equation*}
\xi_{(k)} = \zeta\indices{_{(k)}^{A}} \, \xi_{A} \quad \textnormal{and} \quad \xi^{(k)} = \zeta\indices{^{(k)}_{A}} \xi^{A} \, ,
\end{equation*}
where $\zeta\indices{^{(k)}_{A}} \hspace{-0.07cm} : T\mathbb{C}^2 \rightarrow T\mathbb{C}^2$ are linear mappings with $\zeta\indices{^{(k)}_{A}} \, \zeta\indices{_{(l)}^{A}} = \delta^{(k)}_{(l)}$ and $\zeta\indices{^{(k)}_{A}} \, \zeta\indices{_{(k)}^{B}} = - \delta^{B}_{A}$. Employing the notation $\zeta\indices{_{(0)}^{A}} = o^A$ and $\zeta\indices{_{(1)}^{A}} = \iota^A$, we may express the Newman--Penrose tetrad basis vectors in the dyad spinor representation 
\begin{equation*}
\begin{split}
l^{\mu} & = \sigma\indices{^{\mu}_{A B'}} \, o^A \, \overline{o}^{B'} \\
n^{\mu} & = \sigma\indices{^{\mu}_{A B'}} \, o^A \, \overline{\iota}^{B'}\\
m^{\mu} & = \sigma\indices{^{\mu}_{A B'}} \, \iota^A \, \overline{o}^{B'} \\
\overline{m}^{\mu} & = \sigma\indices{^{\mu}_{A B'}} \, \iota^A \, \overline{\iota}^{B'} \, ,
\end{split}	
\end{equation*}
where
\begin{equation*} 
\sigma\indices{^{\mu}_{A B'}} = \tfrac{1}{\sqrt{2}} \hspace{-0.05cm} \left(\begin{array}{cc}
l^{\mu} & m^{\mu} \\
\overline{m}^{\, \mu} & n^{\mu} \\
\end{array}\right) 
\end{equation*}
are the Infeld--van der Waerden symbols \citep{InfeldVanderWaerden}, a generalization of the usual Pauli matrices.

\section*{Appendix B: Optical Scalars and Sachs Equations} 

\noindent We define Sachs' optical scalars, the key quantities in Penrose's gravito-optical analogy, and their propagation equations within the Newman--Penrose formalism. For this purpose, following \citet[Introduction to Chapter~1.9. and Chapter~1.9.(a)]{ChandraBook}, we first examine the parallel transport of the Newman--Penrose basis vectors $\boldsymbol{l}$, $\boldsymbol{n}$, and $\boldsymbol{m}$ along the direction of $\boldsymbol{l}$. Thus, considering a first-order change in a general basis vector $\boldsymbol{e}_{(a)}$ experiencing an infinitesimal displacement $\boldsymbol{\chi}$, we obtain the relation
\begin{equation} \label{InfDis}
\delta e_{(a) \mu} = [\nabla_{\nu} e_{(a) \mu}] \, \chi^{\nu} = - \gamma_{(a) (b) (c)} \, e\indices{^{(b)}_{\mu}} \, \boldsymbol{\chi}^{(c)} \, .
\end{equation}
This relation yields the change in $\boldsymbol{e}_{(a)}$ per unit displacement along the direction $c$
\begin{equation} \label{Dis}
\delta \boldsymbol{e}_{(a)}(c) = [\nabla_{\nu} \boldsymbol{e}_{(a)}] \, e\indices{_{(c)}^{\nu}} = - \gamma_{(a) (b) (c)} \, \boldsymbol{e}^{(b)} \, ,
\end{equation}
and in particular for the associated changes in $\boldsymbol{l}$, $\boldsymbol{n}$, and $\boldsymbol{m}$ along $\boldsymbol{l}$ 
\begin{align}
\delta \boldsymbol{l}(0) & = 2 \textnormal{Re}(\varepsilon \, \boldsymbol{l} - \overline{\kappa} \boldsymbol{m}) \label{dl} \\[0.2cm]
\delta \boldsymbol{n}(0) & = 2 \textnormal{Re}(- \varepsilon \, \boldsymbol{n} + \pi \, \boldsymbol{m}) \label{dn} \\[0.2cm]
\delta \boldsymbol{m}(0) & = \overline{\pi} \, \boldsymbol{l} - \kappa \, \boldsymbol{n} + 2 i \, \textnormal{Im}(\varepsilon) \, \boldsymbol{m} \, . \label{dm}
\end{align}
Substituting the relation $\delta \boldsymbol{l}(0) = [\nabla_{\nu} \boldsymbol{l}] \, \boldsymbol{l}^{\nu}$, which also can be readily found from Equation (\ref{Dis}) for $a = 0 = c$, into Equation (\ref{dl}), one immediately sees that for $\kappa = 0$ the basis vector $\boldsymbol{l}$ forms a congruence of null geodesics, i.e., a $3$-parameter family of geodesic null curves located in a specific region of spacetime, where each point is intersected by exactly one of these curves. Moreover, for $\varepsilon = 0$, this congruence is affinely parameterized. Without changing these values of $\kappa$ and $\varepsilon$, one can arrange for $\pi$ to vanish as well using a specific class III local Lorentz transformation. \citep[For more details on tetrad transformations see][Chapter~1.8.(g).]{ChandraBook} Then, similarly to $\boldsymbol{l}$, the basis vectors $\boldsymbol{n}$ and $\boldsymbol{m}$ also remain unchanged as they are parallel transported along the general direction of $\boldsymbol{l}$, which is a direct result of Equations (\ref{dn}) and (\ref{dm}). 

From Equation (\ref{Dis}) for $a = 0$, one furthermore obtains the equation
\begin{equation*}
\nabla_{\nu} l_{\mu} = \gamma_{(b) (0) (c)} \, e\indices{^{(b)}_{\mu}} \, e\indices{^{(c)}_{\nu}} \, .
\end{equation*}	
For $\kappa = 0 = \varepsilon$, i.e., for an affinely parameterized geodesic null congruence, this equation can be used to define Sachs' optical scalars \citep*{JordanEhlersSachs,Sachs:1961a}, namely the convergence $\theta$, the rotation $\omega$, and the shear $\sigma$, as
\begin{equation*}
\begin{split}	
\theta & := - \textnormal{Re}(\rho) = \tfrac{1}{2} \, \nabla_{\mu} l^{\mu} \\[0.2cm]
\omega^2 & := \textnormal{Im}^2(\rho) = \tfrac{1}{2} \, \nabla_{[\nu} l_{\mu]} \, \nabla^{\nu} l^{\mu} \\[0.2cm]
|\sigma|^2 & = \tfrac{1}{2} \, \nabla_{(\nu} l_{\mu)} \, \nabla^{\nu} l^{\mu} - \tfrac{1}{4} \, [\nabla_{\mu} l^{\mu}]^2 \, .
\end{split}
\end{equation*}
The geometrical meaning of the optical scalars can be illustrated by drawing a small circle which has its center at a point $p$ on one of the geodesics of the null congruence formed by $\boldsymbol{l}$ and lies in the $2$-plane spanned by $\boldsymbol{m}$ and $\overline{\boldsymbol{m}}$, and then following the geodesics intersecting the circle along the future null direction \citep[cf., e.g.,][Chapter~2.1.3]{GriffithsPodolsky}.
\begin{figure}[t]%
\centering
\includegraphics[width=0.6\columnwidth]{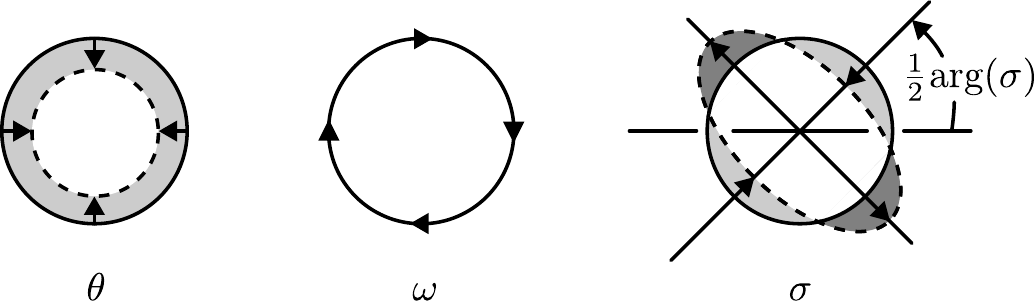}%
\caption[...]
{Geometrical representations of Sachs' optical scalars in terms of the effects of parallel transport of a small perpendicularly oriented circle along the geodesic null congruence.}%
\label{sachs}%
\end{figure}
In doing so, the circle may become expanded or contracted, twisted, and/or sheared into an ellipse, which is measured by $\theta$, $\omega$, and $\sigma$, respectively (see FIG.\ 2). The propagation equations of the optical scalars determining their changes along the geodesic null congruence are given by the corresponding spin coefficient equations \citep[see][Chapter 1.8.(d)]{ChandraBook}
\begin{align}
D \theta & =  \omega^2 - \theta^2 - |\sigma|^2 - \Phi_{0 0} \label{ChangeOS1} \\[0.2cm]
D \omega & = - 2 \theta \omega \label{ChangeOS2} \\[0.2cm]
D \sigma & = - 2 \theta \sigma + \Psi_0 \, , \label{ChangeOS3}
\end{align}
where $D = l^{\mu} \nabla_{\mu}$ and the Ricci and Weyl scalars $\Phi_{0 0}$ and $\Psi_0$ are specified in Equations (\ref{NPRicci}) and (\ref{NPWeyl}). These propagation equations are commonly referred to as the Sachs equations \citep[for example in][]{Perlick}. Finally, the local dyad spinor representations of the optical scalars and the above Ricci and Weyl scalars read
\begin{equation*} 
(- \theta + i \omega) o^A = o_B \, \overline{o}_{C'} \nabla^{A C'} o^B \quad \textnormal{and} \quad \sigma \overline{o}^{A'} = o_B \, o_C \nabla^{C A'} o^B 
\end{equation*}
as well as
\begin{equation*} 
\Phi_{0 0} = \Phi_{A B C' D'} \, o^A \, o^B \, \overline{o}^{C'} \, \overline{o}^{D'} \quad \textnormal{and} \quad \Psi_0 = \Psi_{A B C D} \, o^A \, o^B \, o^C \, o^D \, .
\end{equation*}

\end{appendix}


\end{document}